\documentclass[11pt,a4paper]{article}

\usepackage{jcappub}
\usepackage{slashed}

\usepackage{algorithmic}

\usepackage{multirow}
\usepackage{amssymb}
\usepackage{amsmath}
\usepackage{ulem}
\usepackage{cancel}
\usepackage{verbatim}
\usepackage{caption}
\usepackage{subcaption}
\usepackage[utf8]{inputenc}
\usepackage{xcolor}

\usepackage{bm}
\usepackage{physics}

\usepackage{array}
\newcolumntype{L}[1]{>{\raggedright\let\newline\\\arraybackslash\hspace{0pt}}m{#1}}
\newcolumntype{C}[1]{>{\centering\let\newline\\\arraybackslash\hspace{0pt}}m{#1}}
\newcolumntype{R}[1]{>{\raggedleft\let\newline\\\arraybackslash\hspace{0pt}}m{#1}}

\definecolor{darkred}{rgb}{0.5,0,0}
\definecolor{darkgreen}{rgb}{0,0.5,0}
\definecolor{darkblue}{rgb}{0,0,0.5}

\usepackage{hyperref}
\hypersetup{ colorlinks,
linkcolor=darkblue,
filecolor=darkgreen,
urlcolor=darkred,
citecolor=darkblue }

\newcommand{\inspire}[1]{\href{http://inspirehep.net/search?p=find+J+#1}
 {{\color{black}[{\color{blue} {\small in}SPIRE}]}}}
\newcommand{\book}[1]{\href{http://inspirehep.net/search?p=#1}
 {{\color{black}[{\color{blue} {\small in}SPIRE}]}}}

\newcommand{\inspired}[1]{\href{http://inspirehep.net/search?p=#1}
 {{\color{black}[{\color{blue} {\small in}SPIRE}]}}}

\usepackage{tikz}

\usepackage{txfonts}
\newcommand{\ec}{\eqqcolon}
\newcommand{\ce}{\coloneqq}

\newcommand{\diff}{\mathrm{d}}

\newcommand{\restr}[2]{{
		\left.\kern-\nulldelimiterspace
		#1 
		\vphantom{\big|}\,
		\right|_{#2}
	}}

\begin{document}

\newcommand{\Isabel}[1]{\color{red}{[{\bf Isabel}: #1] }\color{black}}
\newcommand{\Emil}[1]{\color{blue}{[{\bf Emil}: #1] }\color{black}}
\newcommand{\Stefan}[1]{\color{orange}{[{\bf Stefan}: #1] }\color{black}}


\newcommand{\IO}[1]{\textcolor{blue}{Isabel: #1}}

\title{Local clustering of relic neutrinos: Comparison of kinetic field theory and the Vlasov equation} 

\date{\today}

\author[a]{Emil Brinch Holm}
\author[b]{Stefan Zentarra}
\author[c,d]{Isabel M.~Oldengott}

\affiliation[a]{Department of Physics and Astronomy, Aarhus University, DK-8000 Aarhus C, Denmark}

\emailAdd{ioldengott@physik.uni-bielefeld.de}
\affiliation[b]{Institute for Theoretical Physics, ETH Zurich, Wolfgang-Pauli-Str.~27, 8093 Zurich, Switzerland}
\affiliation[c]{Fakultät für Physik, Bielefeld University, D–33501 Bielefeld, Germany}
\affiliation[d]{Centre for Cosmology, Particle Physics and Phenomenology,
Université catholique de Louvain, Louvain-la-Neuve B-1348, Belgium}

\abstract{Gravitational clustering in our cosmic vicinity is expected to lead to an enhancement of the local density of relic neutrinos. We derive expressions for the neutrino density, using a perturbative approach to kinetic field theory and perturbative solutions of the Vlasov equation up to second order. Our work reveals that both formalisms give exactly the same results and can thus be considered equivalent. Numerical evaluation of the local relic neutrino density at first and second order provides some fundamental insights into the frequently applied approach of \textit{linear response} to neutrino clustering (also known as the Gilbert equation). Against the naive expectation, including the second-order contribution does not lead to an improvement of the prediction for the local relic neutrino density but to a dramatic overestimation. This is because perturbation theory breaks down in a momentum-dependent fashion and in particular for densities well below unity.} 


\maketitle   

\flushbottom

\section{Introduction}

The theory of structure formation aims to describe how our Universe evolved from a nearly homogeneous and isotropic state at recombination to the current state that shows a rich variety of structures. On the largest scales the growth of structures is very well described by cosmological linear perturbation theory. However, analytical approaches unfortunately only allow for reliable predictions slightly beyond the linear scale ($k \sim 0.2$ h Mpc$^{-1}$). Here, $N$-body simulations are the standard technique to describe structure formation and manage to reproduce the observed structures in remarkable detail. The high numerical costs of such simulations as well as a possible lack of physical interpretability makes advances in (semi-)analytical approaches nevertheless highly desirable. 

A refreshingly new analytical approach to structure formation is provided by Kinetic Field Theory (KFT)~\cite{Bartelmann2015traj,bartelmann2016KFT,bartelmann2019cosmic,Bartelmann:2020kcx,Konrad:2022fcn,Heisenberg:2022uhb,Pixius:2022hqs}. KFT is based on a path integral formulation of classical mechanics and its main focus has so far been on the calculation of the matter power spectrum, showing promising results. Its formalism, however, may look somewhat complicated and is very different from standard analytical methods. Consequently, KFT has received little attention to date. This motivated us in~\cite{Holm:2023rml} to test the reliability of KFT by applying it to the calculation of the \textit{local density of relic neutrinos}---which constitutes a considerably less complicated cosmological problem than structure formation of cold dark matter (CDM). 

Neutrinos decoupled from the cosmic plasma when the Universe was just about 1 second old. Those remnants from our hot early Universe are called relic neutrinos and form the so-called Cosmic Neutrino Background (C$\nu$B). Observational evidence for the existence of the C$\nu$B currently relies on indirect probes, in terms of measurements of the relativtistic degrees of freedom $N_{\text{eff}}$ through the CMB anisotropy spectrum (e.g.~\cite{Planck:2018vyg}) and through primordial light element abundances as predicted by big bang nucleosynthesis (BBN) (e.g.~\cite{Escudero:2022okz,Pitrou:2018cgg}). A direct detection of the C$\nu$B would be absolutely groundbreaking as it would present the earliest probe of the evolution of our Universe that we currently have. The ambitious PTOLEMY experiment~\cite{Betts:2013uya}  aims to deliver the first-ever direct measurement of the C$\nu$B. While the homogeneous and isotropic background density of relic neutrinos is predicted to be about 56~cm$^{-3}$ per flavour (and about the same for anti-neutrinos), the relic neutrinos' density in our cosmic neighbourhood is expected to be enhanced due to gravitational clustering. Estimating the local overdensity of relic neutrinos is crucial for direct detection experiments since it is proportional to the expected detection rate (for a discussion of current experimental bounds on the overdensity and the sensitivities of future experiments see~\cite{Bauer:2022lri,Brdar:2022kpu,KATRIN:2022kkv,Franklin:2024enc}). Therefore, several works have calculated the local relic neutrino density~\cite{Singh:2002de,Ringwald:2004np,Brandbyge:2010ge,Zhang:2017ljh,deSalas:2017wtt,Mertsch:2019qjv,Zimmer:2023jbb,Elbers:2023mdr}. Current observational bounds on the overdensity are at the order of $10^{11}$ and can be found in  

In~\cite{Holm:2023rml}, we have used KFT at first-order in a perturbation series of the gravitational interaction to calculate the local enhancement of the relic neutrino density and compared it to state-of-the-art results from the back-tracking-method~\cite{Mertsch:2019qjv}. We found extremely good agreement (percent level) between first-order KFT and the results of~\cite{Mertsch:2019qjv} for neutrino masses that are consistent with cosmological neutrino mass bounds ($m_{\nu}\sim 0.1$~eV~\cite{Planck:2018vyg}). For neutrino masses as large as $m_{\nu} \sim 0.3$ eV the agreement was found to degrade to about $50\,\%$. 

We already noted in~\cite{Holm:2023rml} that the analytical expression for the number density from first-order KFT looks similar to the solution of the linearized Vlasov equation, often referred to as \textit{linear response} or the \textit{Gilbert equation} (see e.g.~\cite{Bertschinger:1993xt,Singh:2002de,Ringwald:2004np,Ali-Haimoud:2012fzp,Bird:2018all,Chen:2020kxi,Hotinli:2023scz}). This immediately poses a more fundamental question, concerning a general comparison between KFT and the Vlasov equation. 

The motivation for this works is mainly of conceptual nature, rather than aiming to give the most realistic estimate of the local number density of neutrinos. The two main questions we address in this work are: (i) How does perturbative KFT compare with perturbative solutions to the Vlasov equation? (ii) How much improvement in the prediction of the local relic neutrino density is achieved by including the results from second-order KFT? In order to answer both questions we have extended our work~\cite{Holm:2023rml} to second order in KFT. Somewhat suprisingly, the answer to the second question requires a re-interpretation of our results from~\cite{Holm:2023rml} to some extent. 

The structure of this paper is as follows: In sec.~\ref{sec:State of the art}, we give an overview about different methods that have been applied in order to calculate the local relic neutrino density. We also introduce the so-called $N$-one-body approximation, which is the fundamental reason why the problem at hand is much simpler than CDM structure formation and hence why it so suitable to test KFT. In sec.~\ref{sec:The formalism of kinetic field theory}, we introduce the basic formalism of KFT and in particular show how it simplifies to a formalism of one-particle-quantities (in contrast to a particle ensemble) for the problem at hand. Sec.~\ref{sec:Perturbative KFT} introduces a perturbative approach to KFT and shows the expressions for the neutrino density at first and second order. Sec.~\ref{sec:Comparison to the Vlasov equation} is devoted to the first of the main questions, i.e. here we compare the expressions for the neutrino density from perturbative KFT to perturbative solutions of the Vlasov equation. In sec.~\ref{sec:Local relic neutrino density at second order} we discuss the limitations of perturbative approaches and show the neutrino density at second order in comparison to the results from~\cite{Mertsch:2019qjv}. We conclude finally in sec.~\ref{sec:Conclusions}.


\section{State of the art}
\label{sec:State of the art}

The standard technique to investigate the gravitational clustering of matter is solving  the Vlasov equation (i.e. the collisionless Boltzmann equation) by means of $N$-body simulations. $N$-body simulations are, however, computationally very expensive and embedding neutrinos into CDM simulations is even more challenging (see e.g. \cite{Euclid:2022qde} for a recent summary over the different techniques).

As it was first pointed out in \cite{Singh:2002de} and afterwards applied in \cite{Ringwald:2004np,deSalas:2017wtt,Zhang:2017ljh,Mertsch:2019qjv,Hotinli:2023scz}, calculating the clustering of relic neutrinos in our vicinity does not require full $N$-body simulations, but can be simplified significantly by what we here dub \textit{$N$-one-body approximation}. Within this approximation, the relic neutrinos are treated as test particles in an external gravitational potential governed by CDM and baryons. In other words, it is assumed that 
\begin{itemize}
    \item[i)] the only relevant interaction for neutrinos is gravitation,
    \item[ii)] CDM and baryons evolve independently from neutrinos,
    \item[iii)] neutrinos evolve independently from each other and only feel the gravitational impact of CDM and baryons.
\end{itemize} 
Assumption i) is certainly justified within the standard model of particle physics as neutrinos decoupled long before they started forming structures. Assumptions ii) and iii) are motivated by the fact that relic neutrinos account for at most $1.8\%$ \cite{Planck:2018vyg} of today's matter content of the Universe. On galactic scales this contribution is expected to be even much lower since neutrinos start clustering much later than CDM and baryons. The validity of the $N$-one-body approximation has been confirmed explicitly in \cite{Brandbyge:2010ge}, which we will comment on later in this section. 

In the rest of this section, we give a brief summary of the different techniques that were applied in order to calculate the local density of relic neutrinos (roughly following a chronological order). Note that, except for the full-fledged $N$-body simulations of \cite{Brandbyge:2010ge,Elbers:2023mdr}, all methods are based on the $N$-one-body approximation and this also forms the basis for this work. Due to different assumptions concerning the gravitational potential, a direct one-to-one comparison between the results of the works \cite{Singh:2002de,Ringwald:2004np,Brandbyge:2010ge,Zhang:2017ljh,deSalas:2017wtt,Mertsch:2019qjv,Zimmer:2023jbb,Elbers:2023mdr} is not always possible.  

\paragraph{Linearized Vlasov equation.}
The first technique that has been used in order to calculate the local relic neutrino density has been the solution of the linearised Vlasov equation \cite{Singh:2002de,Ringwald:2004np}. In this work, we go beyond the linear approximation \cite{Singh:2002de,Ringwald:2004np} and also show the results at second order. A comparison between KFT and the solution of the Vlasov equation at first and second order will be the main focus of this work and we dedicate sec.~\ref{sec:Comparison to the Vlasov equation} to a more detailed description of it.

\paragraph{$N$-one-body simulations.}
So called $N$-one-body simulations were first introduced in \cite{Ringwald:2004np}: Instead of simultaneously following the evolution of $N$ particles (neutrinos and CDM) as in conventional $N$-body simulations, the application of the $N$-one-body approximation allows to track one particle at a time and perform $N$ independent simulations. This leads to an enormous reduction of computation time compared to a conventional $N$-body simulations. \cite{deSalas:2017wtt} applied the same method but extended the calculation towards smaller neutrino masses and also improved the modelling of the gravitational potential by taking into account a realistic time evolution of the Milky Way. The point of \cite{Zhang:2017ljh} was mainly to emphasize that it suffices to run one $N$-one-body simulation for an arbitrary neutrino mass and obtain the neutrino overdensity for any neutrino mass by re-weighting. Note that all the works based on the $N$-one-body technique \cite{Ringwald:2004np,deSalas:2017wtt,Zhang:2017ljh} assumed a spherically symmetric gravitational to keep the computation time in a reasonable scale.    

\paragraph{Back-tracking technique.}
The application of the back-tracking method in~\cite{Mertsch:2019qjv} enabled an efficient computation of the local neutrino density including non-spherically symmetric contributions to the gravitational potential (the stellar disk, the Andromeda galaxy, the Virgo cluster etc.). Here, the time direction in the equation of motions is reversed and only those neutrinos ending up \textit{here} and \textit{now} are taken into account. The local overdensity can then be derived by assigning a statistical weight to each trajectory. While the different contributions to the gravitational potential in~\cite{Mertsch:2019qjv} were described in terms of analytical functions, recently the authors of~\cite{Zimmer:2023jbb} refined the calculation by modelling the local gravitational potential on the basis of CDM distributions found in state-of-the-art $N$-body simulations.
In~\cite{Holm:2023rml}, we compared our results from first-order KFT to the results of~\cite{Mertsch:2019qjv}. In this work, we continue using~\cite{Mertsch:2019qjv} as the reference calculation for the comparison with the results from second-order KFT.

\paragraph{$N$-body simulations and constrained simulations.} 
A detailed study of neutrino density profiles in CDM halos was performed in \cite{Brandbyge:2010ge} for a wide range of test halo masses. Here, the neutrino density profiles were calculated both by performing full $N$-body simulations including massive neutrinos and by performing $N$-one-body simulations. It was concluded in \cite{Brandbyge:2010ge} that the results of the $N$-one-body simulations in general agree well with the results of the full $N$-body simulations, which confirms the applicability of the $N$-one-body approximation. Extracting the local neutrino density from the $N$-body simulations in \cite{Brandbyge:2010ge} is of course not directly possible since $N$-body simulations can in general only reproduce local structures in a statistical sense. Very recently, the authors of \cite{Elbers:2023mdr} therefore performed so called constrained simulations in which objects appear at the right positions relative to us.


\section{The formalism of KFT}
\label{sec:The formalism of kinetic field theory}

KFT is based on a path integral formulation of classical mechanics~\cite{Gozzi+1989}. The key idea is to use the very successful formalism of statistical quantum field theory in the context of a classically evolving physical system. It has first been applied to cosmological structure formation in~\cite{bartelmann2016KFT}, and seen further developments and applications over the past years (\cite{bartelmann2019cosmic} for a review). 

The central object in KFT is the generating functional of a particle ensemble. Physical quantities like number densities and higher-order correlation functions can in general be derived by taking functional derivatives of the generating functional. While KFT has so far primarily been applied to structure formation of CDM and in particular the calculation of the matter power spectrum, in~\cite{Holm:2023rml} we applied it for the first time to the local clustering of relic neutrinos. In this section we summarize the formalism of KFT. We explicitly show how the formalism reduces to a formalism of one-particle densities which we have implicitly already assumed in~\cite{Holm:2023rml}.

\subsection{Generating functional}

The generating functional of an ensemble of $N$ neutrinos is given by
\begin{equation}
    Z [\mathbf{J}_y ,\mathbf{y}^{(i)}] = \exp \left( i \int_{z_i}^{z_f} \mathrm{d} z' \, \mathbf{J}_y(z') \cdot \mathbf{y}^{\,s} (z', \mathbf{y}^{(i)}) \right) \,, 
    \label{eq:generating_functional}
\end{equation}
where $\mathbf{J}_y=(\mathbf{J}_x,\mathbf{J}_p)$ is the $6N$-dimensional source function. In order to facilitate the comparison to~\cite{Holm:2023rml} and former works on the relic neutrino density~\cite{Ringwald:2004np,deSalas:2017wtt,Mertsch:2019qjv} we here directly introduce redshift $z$ as our time coordinate. Note that the integral in eq.~\eqref{eq:generating_functional} goes from an initial redshift $z_i$ to the final redshift $z_f$ in the infinite future, $z_f=-1$. The $6N$-dimensional solution to Hamilton's equations of motion is denoted by $\mathbf{y}^{\,s}=(\mathbf{x}^s,\mathbf{p}^s)$ and the $6N$-dimensional initial condition is $\mathbf{y}^{(i)}=(\mathbf{x}^{(i)},\mathbf{p}^{(i)})$. 

When applying the $N$-one-body approximation described in sec.~\ref{sec:State of the art}, it becomes immediately clear that the $6N$-dimensional phase-space trajectory $\mathbf{y}^{\,s}$ can be written as $N$ independent (single-neutrino) trajectories. The generating functional is hence a product of $N$ one-particle generating functionals, i.e.,
\begin{equation}
    Z[\mathbf{J}_y,\mathbf{y}^{(i)}] = \prod_{j=1}^N Z_j \left[ \Vec{J}_{x_j}, \Vec{J}_{p_j}, \Vec{x}^{(i)}_j, \Vec{p}^{(i)}_j \right] \,,
    \label{eq:averaged_Z}
\end{equation}
where we furthermore split the $N$ (6-dimensional) trajectories into their spatial and momentum parts $\Vec{x}^{(i)}_j$ and $\Vec{p}^{(i)}_j$, respectively.

Since our final goal is to calculate the local number density---which is a momentum-independent observable---we may at this point simplify the formalism and directly set the momentum part of the source function to zero, $\mathbf{J}_p=0$ and respectively $\vec{J}_{{p}_j}=0$.  

Let us now take advantage of the fact that we are considering a very large number of neutrinos $N$. In fact, we do not know the exact initial positions and momenta of our $N$ neutrinos (and we do not expect our final result to depend on the exact initial conditions), but we know their initial phase-space distribution. We therefore introduce the generating functional averaged over the initial positions and momenta,  
\begin{equation}
\begin{aligned}
\bar{Z}[\mathbf{J}_x] &= \int \mathrm{d} \mathbf{y}^{(i)} \, P \left( \mathbf{y}^{(i)} \right) \, Z [\mathbf{J}_x,\mathbf{y}^{(i)}] \,,
\end{aligned}
\end{equation}
where $P(\mathbf{y}^{(i)})$ is the probability that the ensemble is characterized by the initial condition $\mathbf{y}^{(i)}$. Since neutrinos decoupled from the cosmic plasma at $T\sim 1$ MeV (when they were still relativistic), their initial momenta follow a relativistic Fermi-Dirac distribution. The initial redshift $z^i$ is then chosen such that the bulk of neutrinos is not yet affected by clustering and thus still matches the homogeneous and isotropic background density. This implies that $P\left( \mathbf{y}^{(i)}\right)$ is independent of the position $\mathbf{x}^{(i)}$ and that the particles are uncorrelated, which implies that $P(\mathbf{y}^{(i)})$ is a product of one-particle probabilities, 
 \begin{equation}
     P \left( \mathbf{y}^{(i)} \right) = \prod_{j=1}^N \frac{1}{N} \frac{1}{(2 \pi)^3} \frac{1}{\exp \left( p^{(i)}_j/T \right)+1} \,, 
 \end{equation}
where we wrote $|\vec{p}^{(i)}_j|= p^{(i)}_j$. Note that the factor of $1/N$ ensures the normalization of $P(\mathbf{y}^{(i)})$.

Together with \eqref{eq:averaged_Z} we find for the averaged generating functional
\begin{equation}
    \bar{Z}[\mathbf{J}_x] = \prod_{j=1}^{N} \frac{1}{N (2 \pi)^{3}} \int \mathrm{d}^3 x_j^{(i)} \, \mathrm{d}^3 \, p_j^{(i)} \frac{1}{\exp \left(p_j^{(i)}/T \right)+1} Z_j \left[  \Vec{J}_{x_j} , \Vec{x}^{(i)}_j, \Vec{p}^{(i)}_j \right] \, ,
    \label{eq:averaged_generating_functional}
\end{equation}
with 
\begin{equation}
    Z_j \left[  \Vec{J}_{x_j} , \Vec{x}^{(i)}_j, \Vec{p}^{(i)}_j \right] = \exp \left( i \int_{z_i}^{z_f} \diff z' \,\Vec{J}_{x_j}(z') \cdot \Vec{x}^{\,s} \left(z',\Vec{x}_j^{(i)},\Vec{p}_j^{(i)}  \right) \right) \, .
    \label{eq:one_particle_Z}
\end{equation}

\subsection{Number density: Reduction to one-particle formalism}
\label{sec:Number density: Reduction to one-particle formalism}

We have seen in the last section how the generating functional factorizes as a product of one-particle generating functionals. In this section, we show explicitly for the example of the number density how the formalism reduces to a formalism of one-particle quantities. This makes the notation applied throughout this work substantially simpler than in most of the KFT literature (e.g. \cite{bartelmann2019cosmic,Heisenberg:2022uhb}). 

The number density of a particle ensemble at position $\vec{x}$ and redshift $z$ is a sum of Dirac delta functions
\begin{equation}
    n(\Vec{x},z) = \sum_{n=1}^N \delta(\Vec{x}-\Vec{x}_n(z)) =   \sum_{n=1}^N \int \frac{ \diff^3 k_n}{(2 \pi)^3} \exp \left( i \Vec{k}_n \cdot \left( \Vec{x}-\Vec{x}_n(z) \right) \right) \, , 
\end{equation}
where we made use of the Fourier representation of the Dirac delta function.
Applying eq.~\eqref{eq:observable_operator} from the appendix we can write the number density in terms of a number density operator (in Fourier space) acting on the generating functional, 

\begin{equation}
    n(\Vec{x},z) = \sum_{n=1}^N \int \frac{ \diff^3 k_n}{(2 \pi)^3} \exp \left(i \Vec{k}_n \cdot \Vec{x} \right) \restr{\exp\left( - i \vec{k}_n \cdot \frac{\delta}{i \delta \vec{J}_{x_n}(z)} \right) \, \bar{Z}\left[ \mathbf{J}_x \right]}{\mathbf{J}_x = 0}\, .
    \label{eq:number_density1}
\end{equation}
Inserting eq.~\eqref{eq:averaged_generating_functional} into eq.~\eqref{eq:number_density1}  we find 
\begin{align}
n(\vec{x},z) 
    &= \restr{ \sum_{n=1}^N  \int \frac{ \diff^3 k_n}{(2 \pi)^3} \exp \left(i \Vec{k}_n \cdot \Vec{x} \right) \exp\left( - i \vec{k}_n \cdot \frac{\delta}{i \delta \vec{J_{x_n}}(z)} \right) \, \prod_{j=1}^{N} \Bar{Z}_j \left[ \Vec{J}_{x_j}\right] }{\vec{J}_{x_j} = 0} \, \\
    &= \sum_{n=1}^N \int \frac{ \diff^3 k_n}{(2 \pi)^3} \exp \left(i \Vec{k}_n \cdot \Vec{x} \right) \left[ \prod_{j \neq n}  \frac{1}{N (2 \pi)^{3}} \int \mathrm{d}^3 x_j^{(i)} \, \mathrm{d}^3 \, p_j^{(i)} \frac{1}{\exp \left(p_j^{(i)}/T \right)+1} \right] \label{eq:product} \\
    & \times \restr{ \left[ \frac{1}{N (2 \pi)^{3}} \! \int \! \mathrm{d}^3 x_n^{(i)} \, \mathrm{d}^3 \, p_n^{(i)} \frac{1}{\exp \left(p_n^{(i)}/T \right)+1}   \exp\left( \! - i \vec{k}_n \! \cdot \! \frac{\delta}{i \delta \vec{J}_{x_n}(z)} \right)  Z_j \left[ \Vec{J}_{x_n}, \Vec{x}^{(i)}_n , \Vec{p}^{(i)}_n \right] \! \right]}{\vec{J}_{x_n} = 0} \,.
    \label{eq:number_Fourier_1}
\end{align}
The product in line~\eqref{eq:product} simply gives a factor $1^{N-1}$ due to normalization. If we now make use of the fact that the ensemble consists of $N$ indistinguishable neutrinos, we can perform the summation over $n$ which gives a factor of $N$. We hence find 
\begin{align}
  n(\vec{x},z) =&\int \frac{ \diff^3 k}{(2 \pi)^3} \exp \left(i \Vec{k}\cdot \Vec{x} \right) \notag \\
  & \hspace{0.5cm} \times \frac{1}{(2 \pi)^{3}} \restr{ \int \mathrm{d}^3 x^{(i)} \, \mathrm{d}^3 p^{(i)} \frac{1}{\exp \left(p^{(i)}/T \right)+1}   \exp\left( - i \vec{k} \cdot \frac{\delta}{i \delta \vec{J_{x}}(z)} \right)  Z \left[ \Vec{J}_{x}, \Vec{x}^{(i)}, \Vec{p}^{(i)} \right]}{\vec{J}_x = 0} \\
  =& \int \frac{ \diff^3 k}{(2 \pi)^3} \exp \left(i \Vec{k}\cdot \Vec{x} \right) \restr{ \int \mathrm{d}^3 x^{(i)} \, \mathrm{d}^3 p^{(i)} f_0(p^{(i)})   \exp\left( - i \vec{k} \cdot \frac{\delta}{i \delta \vec{J_{x}}(z)} \right)  Z \left[ \Vec{J}_{x}, \Vec{x}^{(i)}, \Vec{p}^{(i)} \right]}{\vec{J}_x = 0} \, .
\label{eq:number_density}
\end{align}
This shows how the number density can be written entirely in terms of one-particle quantities.

\subsection{Equations of motion}
\label{sec:Equations of motion}

Let us now work out the equations of motion. Based on our findings in the last section, at this point we can restrict our discussion to the equation of motions of a single neutrino instead of an ensemble of neutrinos.  

\paragraph{Redshift as the time coordinate.}

The Lagrangian for a neutrino with mass $m_{\nu}$, parametrized by cosmic time~$t$ and moving on an expanding background subject to Newtonian interaction with an external gravitational potential $\varphi$ is given by 
\begin{align}
	\mathcal{L}\left(\vec{x},\frac{\diff \vec{x}}{\diff t},t \right) = \frac{m_\nu}{2} a^2(t) \left(\frac{\diff \vec{x}}{\diff t}\right)^2 - m_\nu \varphi\left(\vec{x},t \right) \,.
\end{align}
Note that $\Vec{x}$ as defined in eq.~\eqref{eq:Lagrangian} denotes the \textit{comoving} position $\vec{x}$. 
The peculiar potential $\varphi$ satisfies the comoving Poisson equation
\begin{align}
	\Delta \varphi(\vec{x}, t) = 4\pi G a^2(t) \left( \rho(\vec{x},t) - \bar{\rho}(t) \right) \,.
 \label{eq:Poisson}
\end{align}
In practice, the mean background density~$\bar{\rho}$ is significantly smaller than the halo density profile~$\rho$ and can be ignored to good approximation. A derivation can be found e.g. in the appendix~B of~\cite{Bartelmann2015traj}. 
 
Following former works~\cite{Holm:2023rml,Mertsch:2019qjv,deSalas:2017wtt,Ringwald:2004np} we use redshift~$z$ as our time parameter. This is achieved by transforming the Lagrangian under the condition that the action must be unchanged. We obtain
\begin{align}
	\mathcal{L}\left(\vec{x},\frac{\diff \vec{x}}{\diff z},z\right) &= \frac{\diff t}{\diff z} \left[ \frac{m_\nu}{2} a^2(z) \left(\frac{\diff \vec{x}}{\diff z}\right)^2 \left(\frac{\diff z}{\diff t}\right)^2 - m_\nu \varphi \left(\vec{x},z\right) \right] \\
	&= -\frac{m_\nu}{2} \frac{H(z)}{(1+z)} \left(\frac{\diff \vec{x}}{\diff z}\right)^2 + \frac{m_\nu \varphi\left(\vec{x},z\right)}{(1+z) H(z)} \,,
 \label{eq:Lagrangian}
\end{align}
where we used $\frac{\diff z}{\diff t} = - (1+z) H(z)$, with $H(z)$ denoting the Hubble parameter. Since neutrino clustering mainly happens during the epoch of matter domination, we will assume a two-component Universe, where we have 
\begin{align}
    H(z) = H_0 \sqrt{ \Omega_{m,0} (1 + z)^3 + \Omega_\Lambda } \,.
\end{align}
%
With the conjugate momentum~$\vec{p}$ to the comoving position~$\vec{x}$ \footnote{Note that the momentum as defined in the appendix of ref. \cite{Ringwald:2004np} is not the conjugate momentum.}
\begin{align}
	\vec{p} = \frac{\partial \mathcal{L}}{\partial \left(\frac{\diff \vec{x}}{\diff z}\right)} = - m_\nu \frac{H(z)}{(1+z)} \frac{\diff \vec{x}}{\diff z} \,,
\end{align}
the Hamiltonian is given by 
\begin{align}
	\mathcal{H}\left(\vec{x}, \vec{p}, z\right) = - \frac{1+z}{2 m_\nu H(z)} \vec{p}^2 - \frac{m_\nu \varphi \left(\vec{x},z\right)}{(1+z) H(z)} \,.
\end{align}
This yields the equations of motion
\begin{align}
	\frac{\diff \vec{x}}{\diff z} &= \frac{\partial \mathcal{H}}{\partial \vec{p}} = - \frac{1+z}{m_\nu H(z)} \vec{p} \, , \\
	\frac{\diff \vec{p}}{\diff z} &= -\frac{\partial \mathcal{H}}{\partial \vec{x}} = \frac{m_\nu \frac{\partial \varphi}{\partial \vec{x}} \left(\vec{x},z\right)}{(1+z) H(z)} \, .
	\label{eq:eom}
\end{align}
It is useful to write down the formal solution to these equations of motion. 
This can be achieved by a split into a free ($\varphi=0$) and an interaction part of the solution. The free equation of motion can compactly be written as
\begin{align}
	\frac{\diff}{\diff z} \begin{pmatrix} \vec{x} \\ \vec{p} \end{pmatrix} = \begin{pmatrix} 0 & - \frac{1+z}{m_\nu H(z)} \mathcal{I}_3 \\ 0 & 0 \end{pmatrix} \begin{pmatrix} \vec{x} \\ \vec{p} \end{pmatrix} \ec \mathcal{K} \begin{pmatrix} \vec{x} \\ \vec{p} \end{pmatrix} \,,
 \label{eq:eom_free}
\end{align}
where $\mathcal{I}_3$ is the unity matrix. The solution to eq.~\eqref{eq:eom_free} can be written as 
\begin{align}
	\begin{pmatrix} \vec{x}^{\,s}\left(z; \vec{x}^{(i)}, \vec{p}^{(i)}\right) \\ \vec{p}^{\,s}\left(z ; \vec{x}^{(i)}, \vec{p}^{(i)}\right) \end{pmatrix} = \exp\left( \int_{z_i}^{z} \diff z' \, \mathcal{K}(z') \right) \begin{pmatrix} \vec{x}^{(i)} \\ \vec{p}^{(i)} \end{pmatrix} = \begin{pmatrix} \mathcal{I}_3 & - \int_{z_i}^z \diff z' \, \frac{1+z'}{m_\nu H(z')} \mathcal{I}_3 \\ 0 & \mathcal{I}_3 \end{pmatrix} \begin{pmatrix} \vec{x}^{(i)} \\ \vec{p}^{(i)} \end{pmatrix} \,.
\end{align}
Note that we expanded the exponential in a Taylor series and made use of the fact that the matrix product $\mathcal{K}(z_1) \mathcal{K}(z_2)$ vanishes for any~$z_1, z_2$.
From this we can read off the (retarded) Green's function
\begin{align}
    \mathcal{G}(z_2,z_1) \ce \begin{pmatrix} \, \mathcal{I}_3 & g(z_2,z_1)  \mathcal{I}_3 \\ 0 & \, \mathcal{I}_3 \end{pmatrix} \ce \begin{pmatrix} \mathcal{I}_3 & -\int_{z_1}^{z_2} \diff z' \, \frac{1+z'}{m_\nu H(z')} \mathcal{I}_3 \\ 0 & \mathcal{I}_3 \end{pmatrix}
\end{align}
and be reminded that the redshift integral \textit{always follows the direction of time} (implying $z_1>z_2$). 

We implicitly defined the propagator
\begin{align} \label{eq:defn_propagator}
    g(z_2,z_1) &= -\int_{z_1}^{z_2} \diff z' \, \frac{1+z'}{m_\nu H(z')} \, \\
    &= \left[ \frac{ (1 + z )^2}{2 H_0 m_\nu  \sqrt{\Omega_\Lambda}} \, \, {}_2F_1\left(\frac{1}{2},\frac{2}{3};\frac{5}{3};-\frac{\Omega_{m,0}}{\Omega_\Lambda} (z+1)^3\right) \right]_{z_2}^{z_1} \,, \label{eq:free_propagator}
\end{align}
where $_2F_1$ is the hypergeometric function. Note that the propagator in eq.~\eqref{eq:free_propagator} satisfies ~$g(z_2,z_1) = g(z_2,z_i) - g(z_1,z_i)$ for $z_2 \le z_1 \le z_i$.

For completeness, we write the formal solution to the equations of motion including interactions as
\begin{equation}
\begin{aligned}
\vec{x}^{\,s}(z) &= \vec{x}^{(i)} + g(z,z_i) \vec{p}^{(i)} + m_{\nu} \int_{z_i}^z \mathrm{d}z' \, \frac{g(z,z')}{(1+z') H(z')} \frac{\partial \varphi(\vec{x}^{\,s}(z'))}{\partial \vec{x}} , \\
\vec{p}^{\,s}(z) &= \vec{p}^{(i)} + m_{\nu} \int_{z_i}^z \mathrm{d}z' \, \frac{1}{(1+z') H(z')} \frac{\partial \varphi(\vec{x}^{\,s}(z'))}{\partial \vec{x}} \, .
\label{eq:formal_solution}
\end{aligned}
\end{equation}
Due to its iterative nature the formal solution~\eqref{eq:formal_solution} is of limited practical use.

\paragraph{Super-conformal time as the time coordinate.} 
The main focus of this work is a comparison between the number density from perturbative KFT to perturbative solutions to the Vlasov equation. As the integral solution of the Vlasov equation is conventionally written in terms of super-conformal time $s$  instead of redshift $z$ ($s=\int \mathrm{d}t \, a^{-2}$, where $t$ is cosmic time), we here also show the equations of motion in terms of $s$,
\begin{align}
    \frac{\mathrm{d} \vec{x}}{\mathrm{d}s} &= \frac{\vec{p}}{m_{\nu}}, \\
    \frac{\mathrm{d} \vec{p}}{\mathrm{d}s} &= m_{\nu} a^2(s) \frac{\partial \varphi (\vec{x,s})}{\partial \vec{x}} \, ,
    \label{eq:eom_s}
\end{align}
where we used 
\begin{equation}
    a^2(s) \, \mathrm{d}s = - \frac{\mathrm{dz}}{(1+z) \, H(z)}.
    \label{eq:s_z}
\end{equation}
The propagator is then easily found to be
\begin{equation}
    g(s,s') = \frac{1}{m_{\nu}} (s-s') \, 
\end{equation}
and following the logic from the last paragraph the formal solution to eq.~\eqref{eq:eom_s} is
\begin{equation}
    \begin{aligned}
        \vec{x}^{\,s}(s) &= \vec{x}^{(i)} + \frac{1}{m_{\nu}} (s-s_i) \vec{p}^{(i)} - \int_{s_i}^s \mathrm{d}s' \, a^2(s')\,(s-s') \frac{\partial \varphi(\vec{x}^{\,s}(s'))}{\partial \vec{x}} , \\
\vec{p}^{\,s}(s) &= \vec{p}^{(i)} - m_{\nu} \int_{s_i}^s \mathrm{d}s' \, a^2(s') \, \frac{\partial \varphi(\vec{x}^{\,s}(s'))}{\partial \vec{x}} \, .
\label{eq:formal_solution_s}
    \end{aligned}
\end{equation}

\section{Perturbative KFT}
\label{sec:Perturbative KFT}

We have seen for the example of the neutrino number density in sec.~\ref{sec:Number density: Reduction to one-particle formalism} how within the framework of KFT physical quantities can be derived by differentiation of the generating functional with respect to the source field. However---for most physical systems of interest---deriving expressions that are of direct practical use requires some kind of approximation scheme. 

The two approximations followed so far in the literature are a mean-field approximation~\cite{Bartelmann:2020kcx} and a perturbative approach to KFT~\cite{bartelmann2016KFT,Heisenberg:2022uhb,Pixius:2022hqs}. The mean-field approximation is based on the idea to replace the interactions between particles by an averaged effective force depending on the two-point density correlation function. In this work, we follow the perturbative approach to KFT, and derive the relic neutrino number density at first and second order in the rest of this section.

\subsection{Motivation for a perturbative approach to KFT} 

Let us first try to proceed with the derivation of the number density without applying any approximations. Seeing why and where this attempt fails brings out the necessity for an approximation scheme.  

Performing the functional derivative (see eq.~\eqref{eq:observable_operator} or eq.~\eqref{eq:shift_vector} in appendix~\eqref{app:Formulary}) in eq.~\eqref{eq:number_density} on the generating functional in eq.~\eqref{eq:one_particle_Z} directly gives the following expression for the number density

\begin{align}
      n(\vec{x},z)  &= \int \frac{\diff^3 k}{(2\pi)^3} \exp \left( i \Vec{k} \cdot \Vec{x} \right) \! \int \mathrm{d}^3 x^{(i)} \, \mathrm{d}^3 p^{(i)} \, f_0(p^{(i)}) \, \exp\left( - i \vec{k} \cdot \vec{x}^{\, s}\left( z; \vec{x}^{(i)}, \vec{p}^{(i)} \right) \right) \\
      &= \int \mathrm{d}^3 x^{(i)} \, \mathrm{d}^3 p^{(i)}\, f_0(p^{(i)}) \, \delta \left( \vec{x} - \vec{x}^{\,s}\left( z; \vec{x}^{(i)}, \vec{p}^{(i)} \right) \right) \, ,
      \label{eq:dens_densityDelta}
\end{align}
where we used the Fourier representation of the Dirac delta function. 
The expression in eq.~\eqref{eq:dens_densityDelta} has a transparent interpretation: The density at a position~$\vec{x}$ and redshift~$z$ is given by the particles which end up there at $z$ weighted by the probability of their initial conditions. Eq.~\eqref{eq:dens_densityDelta} is exact but of no direct practical use. Naively, one may now be tempted to simplify eq.~\eqref{eq:dens_densityDelta} by performing the integral over the initial position. As detailed below, \textit{if} the boundary value problem 
\begin{equation}
    \vec{x}^{\,s}\left( z \right) = \vec{x}, \hspace{0.5cm} \vec{p}^{\,s}\left( z_i \right) = \vec{p}^{(i)} \, .
    \label{eq:boundary_value_problem}
\end{equation}
had unique solutions for all $\vec{p}^{(i)}$ this would result in the following expression for the number density
\begin{align}
    n\left( \vec{x}, z \right) = \int \diff^3 p^{(i)} \, f_0(p^{(i)}) \, \left[ \restr{\det\left( \frac{\diff \vec{x}^{\,s}\left( z; \vec{x}^{(i)}, \vec{p}^{(i)} \right)}{\diff \vec{x}^{(i)}} \right)}{\vec{x} = \vec{x}^{\,s}\left( z; \vec{x}^{(i)}, \vec{p}^{(i)} \right)} \right]^{-1} \,.
    \label{eq:number_density_det}
\end{align}

Evaluating the determinant in eq.~\eqref{eq:number_density_det} would now require to find out how the trajectories ending up at $\Vec{x}$ at $z$ with initial momenta $\Vec{p}^{(i)}$ depend on the initial positions $\Vec{x}^{(i)}$. The calculation of the number density in eq.~\eqref{eq:number_density_det} would hence be equivalent to solving the boundary value problem in eq.~\eqref{eq:boundary_value_problem}.  

Indeed, for the trivial case of the free trajecotry (i.e. $\varphi(\Vec{x},z)=0$), one can easily verify that the determinant simply gives a factor of $1$ and the number density in eq.~\eqref{eq:number_density_det} reduces to the standard expression for a homogeneous background of relic neutrinos.

However, for the problem under consideration---and for probably most other applications of interest---the boundary value problem~\eqref{eq:boundary_value_problem} is not unique, i.e. there exist multiple and possibly even infinitely many solutions. Those for example refer to particles performing multiple orbits before arriving at the location $\Vec{x}$ at $z$. The computation of the number density would then require a summation (or in case of infinitely many solutions an integration) over all solutions. If the equations of motions had analytical solutions this would not cause any problem. For our case, however, there exist no analytical solutions and numerically finding all solutions to the boundary value problem~\eqref{eq:boundary_value_problem} is an infeasible task. 
 


\subsection{Perturbative generating functional}
\label{sec:Perturbative Generating functional}

As we saw in the last section, an exact expression for the local relic neutrino number density
can not be obtained due to complications arising from the non-uniqueness of the boundary value problem in eq.~\eqref{eq:boundary_value_problem}. This calls for a perturbative treatment of the generating functional \eqref{eq:averaged_generating_functional}.
Such a perturbative approach has been applied to structure formation in \cite{bartelmann2016KFT,Heisenberg:2022uhb,Pixius:2022hqs}. 

Before we can write down the perturbative expression for the number density, the expression for the generating functional~\eqref{eq:one_particle_Z} has to be brought into a more convenient form. A detailed derivation can be found, e.g., in \cite{Heisenberg:2022uhb}. Here, we directly show the expression that is most useful as our starting point  
\begin{align}
\label{eq:KFT_genFunctional-specific}
	Z\left[\vec{J}_{x}; \vec{x}^{(i)}, \vec{p}^{(i)}\right] &= \exp\left( i \int_{z_i}^{z_f} \diff z' \int_{z_i}^{ z'} \diff z'' \, \frac{m_\nu \, g(z',z'')}{(1+z'') H(z'')} \, \vec{J}_{x}(z') \cdot \frac{\partial \varphi}{\partial \vec{x}} \left(\frac{\delta}{i\,\delta \vec{J}_{x}(z'')},z''\right) \right) \nonumber\\*
	&\qquad \times \exp \left( i \int_{z_i}^{z_f} \diff z' \, \vec{J}_{x}(z') \cdot \left( \vec{x}^{(i)} + g(z',z_i) \vec{p}^{(i)} \right) \right) \,.
\end{align}
The second line in the above expression is identified as the \textit{free generating functional} $Z_0$ (being independent of the gravitational potential) while the first line acts as an interaction operator $\exp \left( i S_I \right)$ on it,
\begin{equation}
    Z\left[\vec{J}_{x}; \vec{x}^{(i)}, \vec{p}^{(i)}\right] = \exp \left( i \hat{S}_I \right) Z_0\left[\vec{J}_{x}; \vec{x}^{(i)}, \vec{p}^{(i)}\right] \,.
    \label{eq:free_generating_functional}
\end{equation}

The idea of perturbative KFT is now to Taylor expand the exponential $\exp \left( i \hat{S}_I \right)$ which leads to a perturbation series for the generating functional and consequently for the number density, as we will see in the next section.

\subsection{Density expressions at first and second order}
\label{sec:Density expressions at first and second order}

Our starting point is the expression for the number density in eq.~\eqref{eq:number_density}. We insert expression eq.~\eqref{eq:free_generating_functional} and write down the series definition for the interaction operator $\exp \left( i \hat{S}_I \right)$. This ultimately leads to a perturbation series for the number density $n(\vec{x},z)$, i.e.,
\begin{equation}
    n(\vec{x},z) = \bar{n}(\vec{x},z) + \delta n^{(1)}(\vec{x},z) + \delta n^{(2)}(\vec{x},z) + \ldots \, .
\end{equation}
Here, the zeroth order is expected to give the homogeneous and isotropic background number density of relic neutrinos $\bar{n}$. Indeed, this is easily verified,
\begin{align}
    \bar{n}(\vec{x},z) &= \int \frac{\diff^3 k}{(2\pi)^3} \, \exp\left( i \vec{k} \cdot \vec{x} \right)  \int \diff^3 x^{(i)} \int \diff^3 p^{(i)} \, f_0(p^{(i)}) \exp\left( - i \vec{k} \cdot \frac{\delta}{i \delta \vec{J}_x(z)} \right) \nonumber\\*
	&\qquad \times \restr{\exp \left( i \int_{z_i}^{z_f} \diff z' \, \vec{J}_x(z') \cdot \left( \vec{x}^{(i)} + g(z',z_i) \vec{p}^{(i)} \right) \right)}{\vec{J}_x = 0} \label{eq:number_bg_1} \\
	    &= \int \diff^3 p^{(i)} \, f_0(p^{(i)}) \int \diff^3 x^{(i)} \int \frac{\diff^3 k}{(2\pi)^3} \exp\left( - i \vec{k} \cdot \left( \vec{x}^{(i)} + g(z,z_i) \vec{p}^{(i)} - \vec{x} \right) \right) \label{eq:number_bg_2} \\
	    &= \int \diff^3 p^{(i)} \, f_0(p^{(i)}) \label{eq:number_bg_3} \,,
\end{align}
which is the standard expression of fermions that decoupled while being relativistic (e.g.~\cite{Kolb:1990vq}). Note that from eq.~\eqref{eq:number_bg_1} to eq.~\eqref{eq:number_bg_2} we applied the functional derivative as in eq.~\eqref{eq:observable_free_functional} (ot the shift vector in eq.~\eqref{eq:shift_vector}) in appendix~\ref{app:Formulary}. In eq.~\eqref{eq:number_bg_2} we applied the Fourier representation of the Dirac delta function. As expected, the explicit dependence on the location~$\Vec{x}$ drops out.

The number density at first order has been derived in the appendix of~\cite{Holm:2023rml} and is given by 
\begin{align}
    \delta n^{(1)}\left( \vec{x}, z \right) &= - \int \diff^3 p^{(i)} \, f_0(p^{(i)})\int_{z_i}^{z} \diff z' \, \frac{m_\nu \, g(z,z')}{(1+z') H(z')} \, \Delta \varphi \left( \vec{x} - g(z,z') \vec{p}^{(i)} ,z'\right) \, .\label{eq:number_density_1st}
\end{align}
As noted in~\cite{Holm:2023rml}, the number density from KFT at first order has a very clear and intuitive interpretation: The neutrino density at the position $\Vec{x}$ is obtained by integrating the external matter density (matter density and the potential are related by the Poisson equation~\eqref{eq:Poisson}) back in time along the free trajectory. 

We derive the number density at second order in appendix~\ref{app:Second order KFT derivation}. Here, we simply show the final expression
\begin{equation}
    \begin{aligned}
    \delta n^{(2)}\left( \vec{x}, z \right) 
    =   \int & \diff^3 p^{(i)} \, f_0(p^{(i)}) \int_{z_i}^{z} \diff z' \, \frac{m_\nu \, g(z,z')}{(1+z') H(z')} \int_{z_i}^{z'} \diff z'' \, \frac{m_\nu}{(1+z'') H(z'')} \\
     \times & \left( g(z,z'') \left[ \frac{\partial}{\partial \vec{x}} \Delta \varphi \left(  \vec{x} - g(z,z'') \vec{p}^{(i)} , z''\right) \right] \cdot \left[ \frac{\partial \varphi}{\partial \vec{x}} \left(  \vec{x} - g(z,z') \vec{p}^{(i)} , z'\right) \right] \right. \\
    & \left. + g(z,z') \left[ \frac{\partial \varphi}{\partial \vec{x}} \left(  \vec{x} - g(z,z'') \vec{p}^{(i)} , z''\right) \right] \cdot \left[ \frac{\partial}{\partial \vec{x}} \Delta \varphi \left(  \vec{x} - g(z,z') \vec{p}^{(i)} , z'\right) \right] \right. \\
    &  \left. + g(z,z'') \left[ \Delta \varphi \left(  \vec{x} - g(z,z'') \vec{p}^{(i)} , z''\right) \right] \, \left[ \Delta \varphi \left(  \vec{x} - g(z,z') \vec{p}^{(i)} , z'\right) \right] \right. \\
    & \left. + g(z,z') \left[ \frac{\partial}{\partial x^{\mu}} \frac{\partial}{\partial x^{\nu}} \varphi \left(  \vec{x} - g(z,z'') \vec{p}^{(i)} , z''\right) \right] \left[ \frac{\partial}{\partial x_{\mu}} \frac{\partial}{\partial x_{\nu}} \varphi \left(  \vec{x} - g(z,z') \vec{p}^{(i)} , z'\right) \right] \right) \, ,
    \end{aligned}
    \label{eq:KFT_2nd_order}
\end{equation}
where we assume summation over $\mu,\nu$ in the last term.

\section{Comparison to the Vlasov equation}
\label{sec:Comparison to the Vlasov equation}

The main focus of this work is a comparison of KFT with the standard approach to structure formation, namely the Vlasov equation (or collisionless Boltzmann equation) which describes the evolution of the phase-space density $f(\Vec{x},\Vec{p})$,
\begin{equation}
    \frac{\partial f }{\partial s} + \frac{\Vec{p}}{m_{\nu}} \cdot \frac{\partial f}{\partial \Vec{x}} - a^2 m_{\nu} \nabla \varphi \cdot \frac{\partial f}{\partial \Vec{p}} = 0 \, .
    \label{eq:Vlasov_equation}
\end{equation}
We remind the reader that $\Vec{x}$ and $\Vec{p}$ are comoving positions and momenta and $s$ is super-conformal time~\eqref{eq:s_z}.  


\subsection{Perturbative KFT in super-conformal time}

In order to facilitate the comparison to the integral solutions of the Vlasov equation~\eqref{eq:Vlasov_equation} let us re-write our KFT expressions for the number densities at first~\eqref{eq:number_density_1st} and second order~\eqref{eq:KFT_2nd_order} in terms of super-conformal time. This is done in a straightforward manner when applying the expressions from sec.~\ref{sec:Equations of motion},
\begin{equation}
    \delta n^{(1)}(\vec{x},s) = \int \mathrm{d}^3 p \, f_0(p) \int_{s_i}^s \mathrm{d} s' \, a^2(s') \, (s-s')  \, \Delta \varphi \left( \vec{x} - \frac{\Vec{p}}{m_{\nu}}\left( s-s' \right), s' \right) \, .
    \label{eq:number_density_s}
\end{equation}
and 
\begin{equation}
    \begin{aligned}
    \delta n^{(2)}\left( \vec{x}, s \right) 
    =  \int & \diff^3 p \, f_0(p) \int_{s_i}^{s} \diff s' \, a^2(s') \, (s-s') \int_{s_i}^{s'} \diff s'' \, a^2(s'') \\
      \times &  \left( (s-s'') \left[ \frac{\partial}{\partial \vec{x}} \Delta \varphi \left(  \vec{x} - \frac{\Vec{p}}{m_{\nu}} (s-s'') , s''\right) \right] \cdot \left[ \frac{\partial \varphi}{\partial \vec{x}} \left(  \vec{x} - \frac{\Vec{p}}{m_{\nu}}(s-s') , s'\right) \right] \right. \\
    & \left. + (s-s') \left[ \frac{\partial \varphi}{\partial \vec{x}} \left(  \vec{x} - \frac{\Vec{p}}{m_{\nu}}(s-s'') , s''\right) \right] \cdot \left[ \frac{\partial}{\partial \vec{x}} \Delta \varphi \left(  \vec{x} - \frac{\Vec{p}}{m_{\nu}}(s-s') , s'\right) \right] \right. \\
    &  \left. + (s-s'') \left[ \Delta \varphi \left(  \vec{x} - \frac{\Vec{p}}{m_{\nu}}(s-s'') , s''\right) \right] \, \left[ \Delta \varphi \left(  \vec{x} - \frac{\Vec{p}}{m_{\nu}} (s-s'), s'\right) \right] \right. \\
    & \left. + (s-s') \left[ \frac{\partial}{\partial x^{\mu}} \frac{\partial}{\partial x^{\nu}} \varphi \left(  \vec{x} - \frac{\Vec{p}}{m_{\nu}}(s-s'') , s''\right) \right] \left[ \frac{\partial}{\partial x_{\mu}} \frac{\partial}{\partial x_{\nu}} \varphi \left(  \vec{x} - \frac{\Vec{p}}{m_{\nu}}(s-s') , s'\right) \right] \right) \, .
    \end{aligned}
    \label{eq:KFT_2nd_order_s}
\end{equation}
We have here skipped the index $^{(i)}$ on the momenta for simplicity.

\subsection{Perturbative solutions to the Vlasov equation}
\label{sec:Perturbative solutions to the Vlasov equation}

The solution to the \textit{linearized} version of the Vlasov equation~\eqref{eq:Vlasov_equation} is well-known either as \textit{Gilbert equation} or \textit{linear response}~\footnote{Usually the term \textit{linear response} refers to the more general case, where the impact of the neutrinos on the gravitational potential is regarded. The formal linear response solution can however of course be applied to simpler case, where the gravitational backreaction of neutrinos is neglected by means of the here applied $N$-one-body approximation.} (see e.g.~\cite{Bertschinger:1993xt,Singh:2002de,Ringwald:2004np,Ali-Haimoud:2012fzp,Bird:2018all,Chen:2020kxi,Hotinli:2023scz}). At higher orders, solutions to the Vlasov equation can be found in~\cite{Fuhrer:2014zka}. The solutions to the Vlasov equations are usually given in Fourier-space, which obscures the direct comparison to our findings from KFT in eqs.~\eqref{eq:number_density_s} and \eqref{eq:KFT_2nd_order_s} to some extent. Let us therefore in this section show in some detail how solutions to the Vlasov equation are obtained in position space. By the end of the section, we will have an answer to the question how perturbative KFT compares to perturbative solutions to the Vlasov equation.

We start by taking the Fourier transform of the Vlasov equation~\eqref{eq:Vlasov_equation},
\begin{equation}
\begin{aligned}
  & \frac{\partial}{\partial s} \tilde{f} (\Vec{k}) + i \frac{\Vec{k}\cdot \Vec{q}}{m_{\nu}} \Bar{f}(\Vec{k}) - i a^2 m_{\nu} \Tilde{\varphi} (\Vec{k}) \, \Vec{k} \cdot \frac{\partial f_0}{\partial \Vec{q}} \\
 & \hspace{3.5cm} - i a^2 m_{\nu} \int \frac{\mathrm{d}^3k_1}{(2\pi)^3} \frac{\mathrm{d}^3 k_2}{(2\pi)^3}  \,(2\pi)^3 \delta \left(\Vec{k}-\Vec{k}_1-\Vec{k}_2 \right) \, \Vec{k}_1 \cdot \frac{\partial \delta \Tilde{f} (\Vec{k}_2)}{\partial \Vec{q}} \, \Tilde{\varphi}(\Vec{k}_1) = 0 \,. 
    \label{eq:Vlasov_Fourier}
\end{aligned}
\end{equation}

As usually in cosmological perturbation theory, we have split the phase-space distribution function into a homogeneous and isotropic background and a perturbation part,
\begin{equation}
    \tilde{f}(\Vec{k},\Vec{p}) = f_0(p) + \delta \tilde{f}(\Vec{k},\Vec{p}) \, .
    \label{eq:pert_ansatz}
\end{equation}

Since the gravitational potential itself is of higher order in perturbation theory, the last term in eq.~\eqref{eq:Vlasov_Fourier} only contributes to orders beyond the linear order. In the main text of this work, we focus on the derivation of the solution of the linearized Vlasov equation while we show the derivation at second order in appendix~\ref{app:Second-order Vlasov derivation}. The linearized Vlasov equation is given by 
\begin{equation}
    \frac{\partial \delta \tilde{f}^{(1)} }{\partial s} + \frac{i \vec{k} \cdot \vec{p}}{m_{\nu}} \delta \tilde{f}^{(1)} - i a^2 m_{\nu} \, \tilde{\varphi} \,  \vec{k} \cdot \frac{\partial f_0 }{\partial \vec{p}} = 0 
    \label{eq:linearized_Vlasov}
\end{equation}
which can be written as
\begin{equation}
    e^{-i \frac{\vec{k}\cdot \vec{p}}{m_{\nu}} s} \frac{\partial}{\partial s} \left[ \delta \tilde{f}^{(1)} \, e^{i \frac{\vec{k}\cdot \vec{p}}{m_{\nu}} s}  \right] - i a^2(s) m_{\nu} \, \tilde{\varphi} \,  \vec{k} \cdot \frac{\partial f_0 }{\partial \vec{p}} = 0 \,,
    \label{eq:trick1}
\end{equation}
and has the solution
\begin{equation}
    \delta \Tilde{f}^{(1)} (\vec{k},\vec{p},s) = i m_{\nu} \vec{k} \cdot \frac{\partial f_0}{\partial \vec{p}} \int_{s_i}^s \mathrm{d}s' \, a^2(s')  \, \tilde{\varphi} (\vec{k}, s') \, e^{i \frac{\vec{k}\cdot \vec{p}}{m_{\nu}} (s'-s)} \, .
    \label{eq:delta_f1_sol}
\end{equation}   

We can now integrate this expression over momentum in order to obtain the number density at linear order,
\begin{align}
  \delta \Tilde{n}^{(1)} (\vec{k},s) &= i m_{\nu} \vec{k} \cdot \int_{s_i}^s \mathrm{d}s' \, a^2(s') \, \tilde{\varphi}(\vec{k},s') \int \mathrm{d}^3p \left[ \frac{\partial}{\partial \vec{p}} \left( f_0 e^{i \frac{\vec{k}\cdot \vec{p}}{m_{\nu}}(s'-s)} \right) -i \frac{\Vec{k}}{m_{\nu}} (s'-s) f_0 \, e^{i \frac{\vec{k}\cdot \vec{p}}{m_{\nu}}(s'-s)}  \right] \label{eq:trick_2}\\
  &= \int_{s_i}^s \mathrm{d}s' a^2(s') \, (s-s') \, \tilde{\varphi}(\vec{k},s') \int \mathrm{d}^3p \, f_0(p) \, e^{i \frac{\vec{k}\cdot \vec{p}}{m_{\nu}}(s'-s)} \, ,
  \label{eq:number_Vlasov_1}
\end{align}
where we applied integration by parts and the divergence theorem $\int_V \mathrm{d}p^3 \, \partial g / \partial \Vec{p} = \oint_{S(V)} \mathrm{d}\Vec{A} \, g = 0$ on the surface $S(V) \rightarrow \infty$. Usually, the last integral in the above expression is now identified as the Fourier transform of the Fermi-Dirac distribution,
\begin{equation}
    F (q) = \int \mathrm{d}^3p \, e^{-i \Vec{p}\cdot\Vec{q}} f_0(p) \, .
    \label{eq:momentum_Fourier}
\end{equation}
giving the well-known linear response solution to the Vlasov equation (e.g.~\cite{Bertschinger:1993xt,Singh:2002de,Ringwald:2004np,Ali-Haimoud:2012fzp,Bird:2018all,Chen:2020kxi,Hotinli:2023scz}),
\begin{equation}
   \delta \tilde{n}^{(1)}(\vec{k},s) = \int_{s_i}^s \mathrm{d} s' \, a^2(s') \, (s-s') \, \tilde{\varphi}(\Vec{k},s') \, F\left( \frac{k (s-s')}{m_{\nu}} \right) \, .
   \label{eq:linear_response}
\end{equation}

In~\cite{Holm:2023rml} we already noted that this expression looks very similar to the expression found in first-order KFT in eq.~\eqref{eq:number_density_s} or~\eqref{eq:number_density_1st}. Remarkably, we now show that both expression are in fact \textit{entirely equivalent}. In order to do so, let us take the Fourier transform of eq.~\eqref{eq:number_Vlasov_1},
\begin{equation}
    \begin{aligned}
         \delta n^{(1)} (\vec{x},s) &= \int \mathrm{d}^3p \, f_0(p) \int_{s_i}^s \mathrm{d}s' \,a^2(s') \, (s-s') \int \frac{ \mathrm{d}^3k}{(2\pi)^3} \, \tilde{\varphi}(\Vec{k},s') \, e^{i \Vec{k}\left(\Vec{x} -\frac{\vec{p}}{m_{\nu}}(s-s') \right)} \, \\
         &= \int \mathrm{d}^3p \,f_0(p) \int_{s_i}^s \mathrm{d}s' \,a^2(s') \, (s-s') \, \Delta \varphi \left( \Vec{x} -\frac{\vec{p}}{m_{\nu}}\left( s-s' \right) ,s' \right) \, .
    \end{aligned}
\end{equation}
Needless to say that the equivalence of both expression can also be shown the other way around: Following the same logic, the well-known linear response expression~\eqref{eq:linear_response} can also be derived by Fourier transformation of the first-order KFT expression eq.~\eqref{eq:number_density_s}. 

To the best of our knowledge, the solution to the linearized Vlasov equation in \textit{position space} has not been presented elsewhere so far. Therefore, the equivalence of the expressions~\eqref{eq:number_density_s} and~\eqref{eq:linear_response} that were found with the two different approaches of perturbative KFT and the Vlasov equation has not been obvious immediately---even though admittedly the two expression looked suspiciously similar to start with. 

The obvious and burning question now arises whether the equivalence between KFT and the Vlasov equation also exists beyond the linear order. To put it briefly, at second order the answer is \textit{yes}. Details can be found in appendix~\ref{app:Second-order Vlasov derivation}~\footnote{The solution to the Vlasov equation at higher orders was derived in~\cite{Fuhrer:2014zka} in Fourier space. Applied to a mixture of DM and neutrinos, the formalism in~\cite{Fuhrer:2014zka} is however somewhat more complicated than required for this work and we hence present the derivation at second order in the appendix, where we also show how the solution in position space is obtained.}. 

\subsection{Discussion}
\label{subsec:Discussion}

We have seen in the previous subsection that perturbative KFT and a perturbative treatment of the Vlasov equation provides the same expressions for the local number density of relic neutrinos at linear and second order. Even though an explicit proof is beyond the scope of this work, our findings strongly suggest \textit{that the equivalence between perturbative KFT and the perturbative Vlasov equation holds at all orders}. 

For the calculation of the local relic neutrino density it can hence be concluded that perturbative KFT does not bring any direct advantage compared to the more conventional approach of solving the Vlasov equation in a perturbative way. We nevertheless believe that the access through KFT has provided insights that have been overlooked so far. In particular, we derived expressions for the number density \textit{in position space} (eqs.~\eqref{eq:number_density_s} and \eqref{eq:KFT_2nd_order_s}) whereas solutions to the Vlasov equation were always given in Fourier space~\cite{Bertschinger:1993xt,Fuhrer:2014zka,Ali-Haimoud:2012fzp,Singh:2002de,Bird:2018all, Ringwald:2004np}. Of course (as explicitly demonstrated in sec.~\ref{sec:Perturbative solutions to the Vlasov equation} and \ref{app:Second order KFT derivation}) the same expressions can be obtained from the Vlasov equation but it has been the access through KFT which has drawn our attention to this. Specially at linear order the expression for the number density in position space~\eqref{eq:number_density_s} has a very intuitive and transparent interpretation---which is not so true for the equivalent expression in Fourier space~\eqref{eq:linear_response}.


\section{Local relic neutrino density at second order}
\label{sec:Local relic neutrino density at second order}

Although the perturbative KFT and Vlasov approaches are equivalent, as shown in sec.~\ref{sec:Comparison to the Vlasov equation}, it remains to be seen how well the perturbation theory applies to the calculation of the local relic neutrino overdensity. In \cite{Holm:2023rml} we found an excellent agreement between the number density from first-order KFT and the back-tracking technique by \cite{Mertsch:2019qjv} for small neutrino masses (consistent with cosmological mass bounds). In this section, we show for the first time by how much the calculation can be improved when the second-order contribution~\eqref{eq:KFT_2nd_order} is included.

The general rule of thumb regarding the linear approximation in eq.~\eqref{eq:linear_response} has been to abandon
it when the overdensity $\delta n^{(1)}_{\nu}/\bar{n}_{\nu}$ exceeds order unity (e.g.~\cite{Bertschinger:1993xt,Ringwald:2004np,Hotinli:2023scz}). Naively, one may therefore expect that up to neutrino masses $m_{\nu} \sim 0.2$ eV (when $\delta n^{(1)}_{\nu}/\bar{n}_{\nu} \sim 1$ as found in~\cite{Holm:2023rml}) perturbation theory is valid and inclusion of the second-order contrubution in eq.~\eqref{eq:KFT_2nd_order} should lead to even better agreement with the back-tracking method~\cite{Mertsch:2019qjv}. However, as we elaborate in this section, this naive expectation turns out to be greatly wrong and in particular the calculation of the second-order contribution to the number density forces us to re-interpret our findings of~\cite{Holm:2023rml} to some extent.

\subsection{Calculation}
\label{sec:calculation}
Rather than obtaining the most realistic result, we are only interested in the comparison of perturbation theory to the back-tracking method, so we employ a simplistic modelling of the gravitational environment of our solar system in the Milky Way as a $2.03\times 10^{12}$ Solar mass spherically symmetric NFW potential~\cite{Navarro:1995iw} representing the central dark matter halo of the Milky Way. This greatly simplifies our calculations, since derivatives of the potential of up to third order enters the second order expression~\eqref{eq:KFT_2nd_order}. Appendix~\ref{app:Gravitational environment} describes the chosen potential and its derivatives in more detail. Reference~\cite{Mertsch:2019qjv} computes the overdensity using the same model (dubbed \textit{NFW} in~\cite{Mertsch:2019qjv}), to which we have matched our parameters in order to facilitate the comparison. The only difference is that the potential of reference~\cite{Mertsch:2019qjv} transitions to a Kepler potential outside of the characteristic radius of the NFW potential; since this substantially complicates the second-order computation and leads to unphysical discontinuities in the higher derivatives of the gravitational potential, we assume the NFW halo to extend to infinity. As we show in appendix~\ref{app:Gravitational environment}, this has only a minor impact on our results. All the results presented in this section are evaluated at a distance of $8.2$ kpc from the NFW center, approximating the distance of our Solar System to the center of the Milky Way.

Using the results~\eqref{eq:number_density_1st} and ~\eqref{eq:KFT_2nd_order}, we compute the relic density as an iterated integral of the form
\begin{align} \label{eq:numerical_1}
    n(\vec{x}) = \int \mathrm{d}^3 p \, f_0 (p) \left[ 1 + \varepsilon^{(1)} (\vec{x}, \vec{p}) +  \varepsilon^{(2)} (\vec{x}, \vec{p}) \right] \, ,
\end{align}
where the first- and second-order contributions are represented as the momentum-dependent corrections $\varepsilon^{(1)} (\vec{x}, \vec{p})$ and $\varepsilon^{(2)} (\vec{x}, \vec{p})$, which are integrals over redshifts that can be read off from equations~\eqref{eq:number_density_1st} and ~\eqref{eq:KFT_2nd_order}. Note that $\varepsilon^{(2)} (\vec{x}, \vec{p})$ does not equal $\varepsilon^{(2)} (\vec{x}, \vec{p})^2$ but contains a term $\propto \varepsilon^{(2)} (\vec{x}, \vec{p})^2$. Appendix~\ref{app:numerical} describes our numerical approach to carrying out the integral in eq.~\eqref{eq:numerical_1}\footnote{Our code is publicly available at \url{https://github.com/EBHolm/KFT-Neutrinos}.}.

\begin{figure}[tb]
    \centering
    \includegraphics[width=\columnwidth]{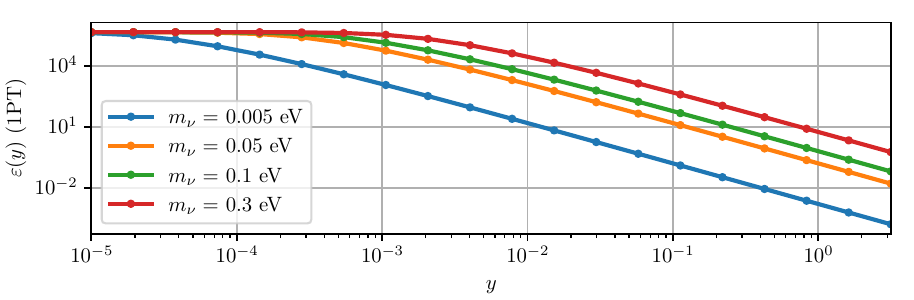}
    \caption{\label{fig:eps_first} First order perturbations to the neutrino distribution function, as a function of their rescaled momenta $y\equiv p/T_0$, for different masses, evaluated at a distance of $x_2=8.2$ kpc from the center of the NFW halo and at a fixed polar angle of the initial momentum $\theta=\pi/2$. The perturbations are monotonically increasing with both momentum and mass.}
\end{figure}

\begin{figure}[tb]
    \centering
    \includegraphics[width=\columnwidth]{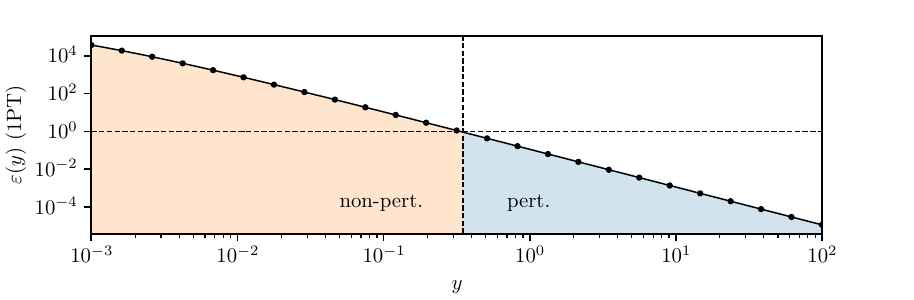}
    \caption{\label{fig:epsilon_fig} First order perturbations to the neutrino distribution function, as a function of the rescaled momentum $y$, for the fixed mass $m=0.05$ eV and polar initial momentum angle $\theta=\pi/2$. The perturbation theory breaks down whenever $\varepsilon(y) \gtrapprox 1$, giving rise to a division of the neutrinos into those that have a small momentum, and therefore are non-perturbative, and those with large momenta that remain perturbative. The momentum cut-off depends widely on parameters like neutrino masses, cluster masses and the initial momentum polar angle $\theta$.}
\end{figure}

It is now crucial to realize that perturbation theory breaks down whenever any of the momentum-dependent corrections\footnote{In the following, we suppress the obvious dependence on $\vec{x}$.} $\varepsilon^{(i)} (\vec{p})$ are of order unity or above. This becomes most clear from the derivation of eq.~\eqref{eq:numerical_1} starting with the Vlasov equation in sec.~\ref{sec:Perturbative solutions to the Vlasov equation}, in particular the perturbation ansatz in eq.~\eqref{eq:pert_ansatz}. In other words, when perturbation theory breaks down is in general a momentum-dependent question. This has also been noted in reference~\cite{Ringwald:2004np}.
Figure~\ref{fig:eps_first} shows the values of $\varepsilon^{(1)} (y)$ for different neutrino masses at fixed positions and momentum directions, as functions of the rescaled momenta $y\equiv p/T_0$. In all cases, they are monotonically decreasing functions of $y$, the normalised momentum: At large $y$, they are much smaller than unity, but at small $y$, they become large and break the applicability of perturbation theory. The interpretation of this is that for any assumed neutrino mass, there will always exist neutrinos with sufficiently small momenta that they are non-perturbative. This is illustrated schematically in figure~\ref{fig:epsilon_fig}. 

How many of such neutrinos there are is dictated by their initial Fermi-Dirac distribution, and for small enough neutrino masses, they will have a negligible impact on the overdensity. However, this balance is also affected by the assumed mass of the gravitational cluster, and for the mass of the Milky Way, we find that there is always a substantial amount of non-perturbative neutrinos, meaning that the overdensity often becomes unphysical. In conclusion, therefore, the Milky Way is simply too massive for perturbative methods to apply directly to the calculation of relic neutrino number overdensities\footnote{Note that higher-order moments of the distribution function, such as energy density or pressure, will be less sensitive to the non-perturbative neutrinos due to the weighting of $y$ to a higher-order power in the momentum integral.}. 

Nonetheless, the perturbative methods can still give approximate lower bounds on the overdensity by excluding the contribution of the non-perturbative neutrinos. In practice, we accomplish this by only integrating the momentum above $y_\mathrm{cut}$, where $y_\mathrm{cut}$ is a momentum cut-off that excludes the non-perturbative regime\footnote{Other approaches, such as~\cite{deSalas:2017wtt,Mertsch:2019qjv}, also assumed momentum cuts on the order of $y \sim 10^{-2}$, but since these are non-perturbative methods we should not compare our momentum cut-offs to theirs.}. Suitable values of $y_\mathrm{cut}$ can be inferred by requiring that the numerical value of the first order contribution $\varepsilon^{(1)} (y_\mathrm{cut})$ is smaller than some threshold $\varepsilon_\mathrm{cut}$, which must be smaller than order unity\footnote{If the neutrinos at $y=y_\mathrm{cut}$ are perturbative at first order, they will also be perturbative at second order, so we only need to put the requirement on the first order contribution $\varepsilon^{(1)} (y_\mathrm{cut})$.}. Numerically, we invert the equation $\varepsilon^{(1)} (y_\mathrm{cut}, \theta) = \varepsilon_\mathrm{cut}$, for each fixed $\theta$ in the quadrature, to find $y_\mathrm{cut}$ using bracketing and subsequent bisection~\cite{Press2007}.

\begin{figure}[tb]
    \centering
    \includegraphics[width=\columnwidth]{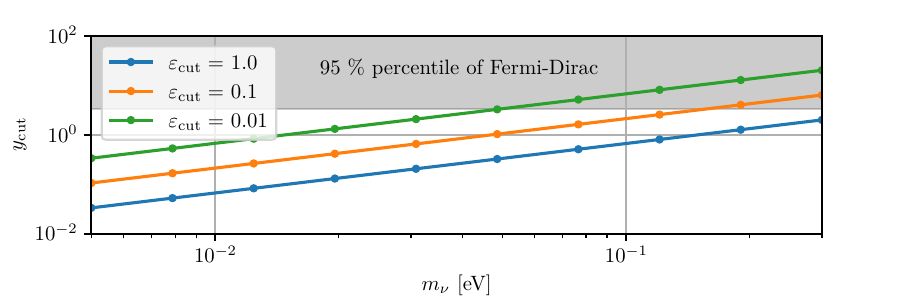}
    \caption{\label{fig:y_cut} Momentum cut-off $y_\mathrm{cut}$, for different masses and absolute cut-offs $\varepsilon_\mathrm{cut}$, below which the neutrinos are non-perturbative. At large masses, a substantial part of the neutrinos become non-perturbative. Here, the initial momentum polar angle is fixed at $\theta=\pi/2$.}
\end{figure}
Figure~\ref{fig:y_cut} illustrates the values of the momentum cut-off $y_\mathrm{cut}$ thus found, for different masses and different tolerances $\varepsilon_\mathrm{cut}$. Evidently, a progressively larger fraction of the neutrinos are non-perturbative with increasing mass. Noting that the Fermi-Dirac distribution already reaches its 95 \% percentile at around $y=3.3$, most of the neutrinos are already non-perturbative at quite small masses. 

This may leave out a potentially large contribution to the overdensity, and is therefore only a lower bound. Since the correct solution would require knowledge of the underlying true, non-linear value of $\varepsilon (y)$, it is difficult to estimate the error thus made. A crude, order of magnitude estimate of the error can be obtained by assuming a fixed value of $\varepsilon^{(i)} (y) = \varepsilon^{(i)} (y_\mathrm{cut})$ for all $y < y_\mathrm{cut}$. The resulting density can then be interpreted as the density if all neutrinos were perturbative. While somewhat arbitrary, this gives a rough indication of what one can expect the error of the perturbative result to be. 

\begin{figure}[tb]
    \centering
    \includegraphics[width=\columnwidth]{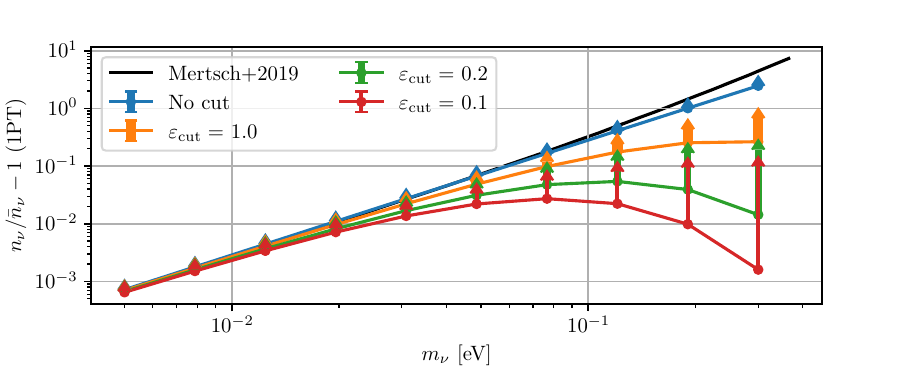}
    \caption{\label{fig:masses_first_order} Estimates of the relic neutrino overdensities at Earth from the non-linear back-tracking method of reference~\cite{Mertsch:2019qjv} (black) and from first order perturbation theory with different values of the cut-off scale $\varepsilon_\mathrm{cut}$. The latter determines how aggressively one excludes the gradually more non-perturbative neutrinos with low momenta; consequently, including only the perturbative neutrinos results in underestimations of the non-linear result. The upwards-facing arrows represent crude order-of-magnitude error estimates described in the text.}
\end{figure}
This still leaves open the choice of a particular cut-off value $\varepsilon_\mathrm{cut}$. Figure~\ref{fig:masses_first_order} shows the first-order overdensity predictions as a function of the neutrino mass, for different values of the cut-off. The black line corresponds to the back-tracking results of reference~\cite{Mertsch:2019qjv} which we take as the gold standard in this comparison. The error estimate described above is indicated by upwards-facing arrows. While all lines converge to the correct result at small masses due to all neutrinos becoming perturbative, they progressively underpredict the true value at masses above $\approx 10$ meV, even when taking the error estimate into account. Interestingly, the prediction with no momentum cut-off, also reported in our earlier paper~\cite{Holm:2023rml}, gives the best approximation of the true result. We discuss this observation in the next section~\ref{subsec:discussion_6}. 
Nonetheless, in the rest of the paper, we choose $\varepsilon_\mathrm{cut}=1.0$ to avoid contributions from regimes where the perturbation theory is explicitly invalid.

\begin{figure}[tb]
    \centering
    \includegraphics[width=\columnwidth]{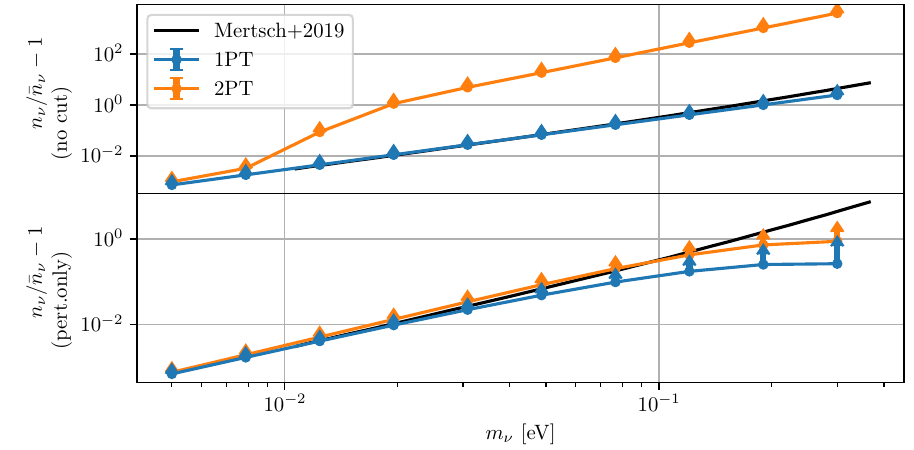}
    \caption{\label{fig:masses_with_second} Estimates of the relic neutrino overdensities from the back-tracking method of reference~\cite{Mertsch:2019qjv} (black) and first- (blue) and second- (yellow) order perturbation theory as described in this paper. In the top panel, no momentum cut-off has been made, and the contribution from the non-perturbative neutrinos make the second-order result diverge. In the bottom panel, only the neutrinos with perturbative distribution function corrections are included. The upwards-facing arrows denote the rough error estimate described in the text.}
\end{figure}
Hence, figure~\ref{fig:masses_with_second} presents the final estimates of the relic neutrino overdensity from first- and second-order perturbation theory as a function of the neutrino mass. Again, the black lines denote the (for the sake of comparison) reference calculation of~\cite{Mertsch:2019qjv}. In the top panel, no momentum cut-off is made in the perturbation theory. While the first-order result agrees very well with the reference calculation the second order overestimates the reference for $m_{\nu}\gtrapprox 10$ meV by several orders of magnitude due to the contribution of non-perturbative neutrinos. The bottom panel shows the same, but only taking the contribution from neutrinos with perturbative momenta into account. Evidently, neither the first- nor the second-order overestimate the reference result\footnote{The slight $\approx 10 \%$ overestimation at small masses is due to small differences in modelling of the gravitational potential, such as the transition to a Kepler potential made in~\cite{Mertsch:2019qjv}.}. As generically expected from any perturbation series, the inclusion of the second-order contribution leads to better agreement with the correct result. 

\subsection{Discussion}
\label{subsec:discussion_6}
 
Let us now turn the discussion to the final question how well perturbation theory applies to the calculation of the local neutrino density. 

\begin{figure}[tb]
    \centering
    \includegraphics[width=\columnwidth]{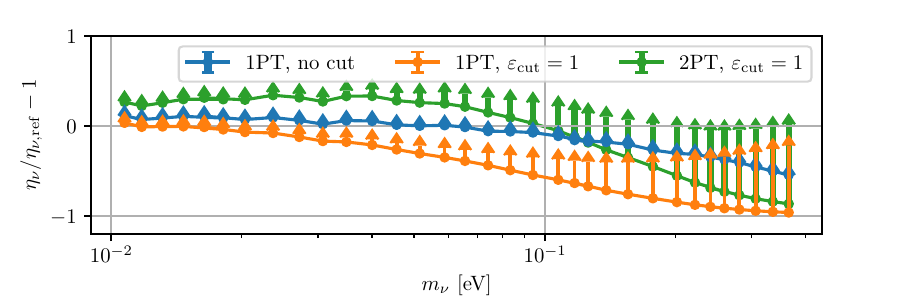}
    \caption{\label{fig:masses_diffs_single} Relative differences between the relic neutrino overdensities $\eta_\nu \equiv n_\nu / \bar{n}_\nu - 1$ as calculated with the back-tracking method~\cite{Mertsch:2019qjv} and first-order perturbation theory without applying any momentum cut-off (blue), first-order perturbation theory with $\epsilon_{\text{cut}}=1.0$ (orange), and second-order perturbation theory with $\epsilon_{\text{cut}}=1.0$ (green). }
\end{figure}

Fig.~\ref{fig:masses_diffs_single} shows the relative difference between the overdensities $\eta_\nu \equiv n_\nu / \bar{n}_\nu - 1$ from the back-tracking method~\cite{Mertsch:2019qjv} and i) first-order perturbation theory without applying any momentum cut-off (i.e. the results from~\cite{Holm:2023rml}), ii) first-order perturbation theory with $\epsilon_{\text{cut}}=1.0$, and iii) second-order perturbation theory with $\epsilon_{\text{cut}}=1.0$. 

We observe that, after all, the best agreement to the back-tracking method over the full neutrino mass range is observed \textit{from first-order perturbation theory when no momentum cut-off is applied} (i.e. the results from~\cite{Holm:2023rml}). This may come as a surprise since we have seen in the last section that this calculation contains a significant contribution from momenta where perturbation theory is not valid. It is currently unknown whether this is an accident. The most plausible reason to us is the combined effect of the inherent underestimation due to only taking into account the first-order contribution and the overestimation due to taking into account the non-perturbative momentum regime.

When the non-perturbative momenta are cut off, the agreement between first-order perturbation theory and the back-tracking method degrades tremendously. Fortunately, the inclusion of the second-order contribution leads to significant improvement at neutrino masses $m_{\nu}>0.1$ eV, especially when the error estimate is taken into account. However, at lower neutrino masses the second-order results does not converge to the reference and in particular it even overestimates the overdensity, which seems very unphysical. We believe that this behaviour is caused by differences in the modelling of the gravitational potential, in particular the transition to a Kepler potential made in~\cite{Mertsch:2019qjv} (as discussed in more detail in app.~\ref{app:Gravitational environment})\footnote{Different from~\cite{Holm:2023rml}, we here also omit the Kepler transition at the first order, which is why the convergence at low momenta looks somewhat less smooth than in~\cite{Holm:2023rml}.}. We want to emphasize again that the inclusion of this transition would itself be unphysical as the higher derivatives of the gravitational potential (required for the second-order contribution~\eqref{eq:KFT_2nd_order}) would become discontinuous---a problem that neither arises for the back-tracking method nor first-order perturbation theory. Needless to say that reproducing the results of~\cite{Mertsch:2019qjv} without the transition to the Kepler potential is entirely out of the scope of this work. Finally, it is also not entirely clear how much impact this different modelling of the gravitational potential has on larger masses $m_{\nu}>0.1$ eV. Unfortunately, this makes a final conclusion about the reliability of second-order perturbation theory very hard at this point.


\section{Conclusions}
\label{sec:Conclusions}

In this work, we have dealt with a comparison of perturbative KFT and perturbative solutions of the Vlasov equation. We also addressed questions concerning the general limitations of perturbation theory, which is particularly interesting in light of the frequently applied linear response approach to describe the clustering behaviour of neutrinos \cite{Bertschinger:1993xt,Singh:2002de,Ringwald:2004np,Ali-Haimoud:2012fzp,Bird:2018all,Chen:2020kxi,Hotinli:2023scz}.
We thereby focussed on the calculation of the number density of relic neutrinos in our cosmic vicinity. As discussed in sec.~\ref{sec:State of the art} and \ref{sec:The formalism of kinetic field theory}, the application of the $N$-one-body approximation simplifies the problem significantly since the  formalism reduces to a formalism of single particles in an external gravitational potential. This makes the problem particularly suited to test the reliability of KFT. At the same time, estimating the local density of relic neutrinos is crucial for future experiments aiming to deliver the first direct detection of the relic neutrino background.  

We extended our former work~\cite{Holm:2023rml} to second order in a perturbation series of the gravitational interaction for KFT. We also derived the integral solution of the Vlasov equation at second order in standard (Eulerian) perturbation theory. Our work shows that perturbative KFT gives exactly the same results as the Vlasov equation. While we have shown this explicitly at first and second order, our work is strongly suggestive that \textit{this equivalence holds at all orders.} On very general grounds, this equivalence can be expected (despite not being formally obvious): While the two formalisms look extremely different, both KFT and the Vlasov equation (also known as single-particle Liouville equation or collisionless Boltzmann equation) are purely based on classical mechanics.  
This now also raises the question concerning a general comparison between KFT and standard cosmological perturbation theory. Clearly, we cannot provide an entirely generic and final answer to this question, given that the problem addressed in this work justified a couple of assumptions to be made. To be explicit, we have demonstrated the equivalence between perturbative KFT and Eulerian perturbation theory under the assumption of an \textit{initially uncorrelated homogeneous distribution of test particles in an external matter field}. In the context of CDM clustering the same set of assumptions would clearly be invalid. Here, the backreaction of CDM on the the gravitational potential makes the analagous integral solutions~\eqref{eq:number_density_1st} and \eqref{eq:KFT_2nd_order} inconvenient due to their recursiveness. Instead, the standard procedure in cosmological perturbation theory is to integrate the Vlasov equation over momentum in order to obtain a hierarchy of coupled differential equations for the density contrast, bulk velocity, anisotropic stress and so on. This hierarchy of equations is then usually cut at the level of the anisotropic stress, which causes standard cosmological perturbation theory to break down once shell-crossing occurs. The alternative approach through KFT, either by a perturbation series or a mean-field approach \cite{bartelmann2016KFT,Pixius:2022hqs}, may in fact turn out to be advantageous in this case. On the other hand, our work has shown that already for masses $m \gtrsim 0.2$~eV second-order KFT becomes unreliable. For the case of CDM structure formation this would then imply that contributions up to very high order have to be taken into account.

The second objective of this work has been the calculation of the  second-order contribution to the relic neutrino density. 
We compared our results to those from the back-tracking method~\cite{Mertsch:2019qjv} and concentrated on a spherically symmetric NFW profile for the gravitational potential. When no momentum cut-off for the low-energetic neutrinos is applied (as has been the case in our previous work~\cite{Holm:2023rml}), the second-order contribution to the number density greatly overestimates the number density of neutrinos. As we have shown in this work, this is due to the fact that a large part of the low-energetic neutrinos become non-perturbative. The prevalence of non-perturbative neutrinos is strongly dependent on the parameters of the problem, most importantly the assumed neutrino mass and the virial mass of the gravitational profile. In other words, given our modelling of the gravitational environment, the Milky Way is too heavy for the perturbation theory to be adequately applicable to this problem. Furthermore, we believe that the good agreement between the results from first-order KFT and the back-tracking technique found in~\cite{Holm:2023rml} is caused by an accidental cancellation of two effects: underestimating the result by only taking into account the linear order and at the same time overestimating the result by taking into account the non-perturbative momentum regime. Note that due to the equivalence of perturbative KFT and perturbative solutions of the Vlasov equation the same also applies to the frequently used approach of \textit{linear response} to model the clustering behaviour of neutrinos (e.g.~\cite{Bertschinger:1993xt,Singh:2002de,Ringwald:2004np,Ali-Haimoud:2012fzp,Bird:2018all,Chen:2020kxi,Hotinli:2023scz}). When applying a momentum cut-off for the non-perturbative momenta, the agreement between first-order KFT and the back-tracking method is degraded significantly---but significantly improved again when taking into account the second-order contribution (as expected). Due to differences in the modelling of the gravitational potential between our work and ref.~\cite{Mertsch:2019qjv} at this point it is unfortunately not possible to draw a final conclusion on how well second-order perturbation theory describes the clustering behaviour of neutrinos. 

To sum up, given that currently linear response or first-order KFT shows the best agreement to the back-tracking method, from a pragmatic point of view there is nothing that speaks against applying it to describe the gravitational clustering of neutrinos. However, on general grounds, the validity of linear response should not be confused with the question on when perturbation theory breaks down. We have shown in this work that the validity of perturbation theory depends not only on the neutrino mass but also on the momentum and the gravitational potential (in our case the virial mass of the DM halo).

\acknowledgments
We thank Matthias Bartelmann, Steen Hannestad, Thomas Tram and Yvonne Y. Y. Wong for valuable discussions. IMO acknowledges support by Fonds de la recherche scientifique (FRS-FNRS). EBH was supported by a research grant (29337) from VILLUM FONDEN.

\appendix

\section{Details on KFT}
\label{sec:Details on kinetic field theory}

\subsection{Formulary}
\label{app:Formulary}
In this appendix, we show some useful relations that are needed to understand how KFT can be used in order to calculate physical observables, such as in our case the relic neutrino overdensity. All of these equations can as well be found e.g. in ~\cite{bartelmann2019cosmic} (plus references within) and~\cite{Bartelmann:2020kcx,Konrad:2022fcn,Heisenberg:2022uhb,Pixius:2022hqs}. However, due to the reduction to a formalism of one-particle quantities, the notation of this appendix is much simpler than in the rest of the KFT literature. 

The most basic and most important relation is of course the functional derivative which obeys the basic axiom 
\begin{equation}
    \frac{\delta }{\delta \vec{J}_x (z)} \Vec{J}_x(z') = \delta(z-z') \hspace{1cm} \text{or} \hspace{1cm} \frac{\delta }{\delta \vec{J}_x (z)} \int \diff z' \, \Vec{J}_x (z') \cdot \Vec{x}(z') = \Vec{x}(z) \, . 
    \label{eq:Functional_derivative}
\end{equation}
Applying the ordinary rules for derivatives, the average position at redshift $z$ is then given by 
\begin{equation}
\langle \Vec{x}(z) \rangle = - i \frac{\delta}{\delta \Vec{J}_{x}(z)} \restr{\bar{Z} \left[ \Vec{J}_x \right]}{\Vec{J}_x=0} \, . 
\end{equation}
More generally, any observable $\mathcal{O}$ that can be expressed as an analytical function of the position $\Vec{x}$ can be derived upon differentiation of the generating functional with respect to the source field,
\begin{equation}
    \mathcal{O}(\Vec{x}) = \restr{\mathcal{\hat{O}} \left( \frac{\delta}{i \delta \Vec{J}_x (z)} \right) \Bar{Z} \left[ \Vec{J}_x \right] }{\Vec{J}_x=0} \, .
    \label{eq:observable_operator}
\end{equation}
Here, the use of the $\, \hat{} \,$ symbol should emphasize that the corresponding function is understood as an operator, i.e. an object acting on the generating functional.

When applied on the free generating functional $Z_0$ (defined as in eq.~\eqref{eq:KFT_genFunctional-specific} and eq.~\eqref{eq:free_generating_functional}), we find  
\begin{equation}
       \mathcal{\hat{O}}  \restr{\left( \frac{\delta}{i \delta \vec{J}_x (z)} \right) Z_0 \left[ \vec{J}_x, x^{(i)}, p^{(i)} \right]}{\vec{J}_x = 0} = \mathcal{O} \left( \vec{x}^{(i)} + g(z,z_i) \vec{p}^{(i)} \right) \, .
       \label{eq:observable_free_functional}
    \end{equation}  

It is also very useful to note that the number operator $\exp \left( -i \Vec{k}\cdot \frac{\delta}{i \delta \Vec{J}(z)} \right)$ applied on the source function shifts the source function according to (see eq. (28) in \cite{bartelmann2019cosmic})
\begin{equation}
    \vec{J}_x(z') \rightarrow \Vec{J}_x(z) - \delta(z-z') \Vec{k} \, .
    \label{eq:shift_vector}
\end{equation}

\subsection{Second order KFT derivation}
\label{app:Second order KFT derivation}

This appendix contains the derivation of the second order contribution to the relic neutrino number density from KFT in eq.~\eqref{eq:KFT_2nd_order}. The starting point is again eq.~\eqref{eq:number_density1}, where we now insert the second-order contribution of the interaction operator~\eqref{eq:KFT_genFunctional-specific},

\begingroup
\allowdisplaybreaks
\begin{align}
    \delta n^{(2)}\left( \vec{x}, z \right) 
        &= \int \frac{\diff^3 k}{(2\pi)^3} \, \exp\left( i \vec{k} \cdot \vec{x} \right) \int \diff^3 x^{(i)} \int \diff^3 p^{(i)} \, f_0(p^{(i)}) \exp\left( - i \vec{k} \cdot \frac{\delta}{i \delta \vec{J}_x(z)} \right) \\*
    &\qquad \times \frac12 \left( i \int_{z_i}^{z_f} \diff z_1' \int_{z_i}^{z_1'} \diff z_1'' \, \frac{m_\nu \, g(z_1',z_1'')}{(1+z_1'') H(z_1'')} \, \vec{J}_x(z_1') \cdot \frac{\partial \varphi}{\partial \vec{x}} \left(\frac{\delta}{i\,\delta \vec{J}_x(z_1'')},z_1''\right) \right) \\*
    &\qquad \times \left( i \int_{z_i}^{z_f} \diff z_2' \int_{z_i}^{z_2'} \diff z_2'' \, \frac{m_\nu \, g(z_2',z_2'')}{(1+z_2'') H(z_2'')} \, \vec{J}_x(z_2') \cdot \frac{\partial \varphi}{\partial \vec{x}} \left(\frac{\delta}{i\,\delta \vec{J}_x(z_2'')},z_2''\right) \right) \\*
    &\qquad \times \restr{\exp \left( i \int_{z_i}^{z_f} \diff z' \, \vec{J}_x(z') \cdot \left( \vec{x}^{(i)} + g(z',z_i) \vec{p}^{(i)} \right) \right)}{\vec{J}_x = 0} \\
        &= \frac12 \int \frac{\diff^3 k}{(2\pi)^3} \, \exp\left( i \vec{k} \cdot \vec{x} \right) \int \diff^3 x^{(i)} \int \diff^3 p^{(i)} \, f_0(p^{(i)}) \, \exp\left( - i \vec{k} \cdot \frac{\delta}{i \delta \vec{J}_x(z)} \right) \\*
    &\qquad \times \left( \int_{z_i}^{z_f} \diff z_1' \int_{z_i}^{z_1'} \diff z_1'' \, \frac{m_\nu \, g(z_1',z_1'')}{(1+z_1'') H(z_1'')} \, i\vec{J}_x(z_1') \cdot \frac{\partial \varphi}{\partial \vec{x}} \left(\frac{\delta}{i\,\delta \vec{J}_x(z_1'')},z_1''\right) \right) \label{eq:term1} \\*
    &\qquad \times \left( \int_{z_i}^{z_f} \diff z_2' \int_{z_i}^{z_2'} \diff z_2'' \, \frac{m_\nu \, g(z_2',z_2'')}{(1+z_2'') H(z_2'')} \, i\vec{J}_x(z_2') \cdot \frac{\partial \varphi}{\partial \vec{x}} \left(\left( \vec{x}^{(i)} + g(z_2'',z_i) \vec{p}^{(i)} \right),z_2''\right) \right) \label{eq:term2} \\*
    &\qquad \times \restr{\exp \left( i \int_{z_i}^{z_f} \diff z' \, \vec{J}_x(z') \cdot \left( \vec{x}^{(i)} + g(z',z_i) \vec{p}^{(i)} \right) \right)}{\vec{J}_x = 0}\, , 
\end{align}
\endgroup
where we applied eq.~\eqref{eq:observable_free_functional} in eq.~\eqref{eq:term2}.

Next we essentially want to commute the two terms in eq.~\eqref{eq:term1} and~\eqref{eq:term2} such that we can act with the functional derivative in eq.~\eqref{eq:term1} also on the free generating functional in the last line. In order to do so, we need to move the term~\eqref{eq:term1} past the factor $\vec{J}_x(z_2')$ in term \eqref{eq:term2}. This is done in terms of a product rule (or a commutator term, depending on how you want to interpret it), hence we get two terms. The first is simply the one where we have exchanged the position of the terms \eqref{eq:term1} and \eqref{eq:term2}, but the second is more complicated. What happens is essentially 
\begin{equation}
    f(\partial_J) ( J \cdot g ) = \sum\limits_{n=1} f^{(n)}(0) \frac{\partial_J^n}{n!} ( J \cdot g ) = \sum\limits_{n=1} f^{(n)}(0) \frac{\partial_J^{n-1}}{(n-1)!} \cdot g = g \cdot \nabla f(\partial_J) \, .
\end{equation} 
Here we used that $f(0) = 0$~\footnote{As a further remark on $f(0) = 0$: Note first that the development point for the Taylor expansion is indeed around $0$, since each term has $(\partial_J - 0)^n$. Next, it is $f(0) \equiv \frac{\partial \varphi}{\partial \vec{x}}(0) = 0$, i.e., the force due to the external potential vanishes in the center of the halo. This is a convention which avoids having to deal with certain factors of $\infty$ elsewhere, and is done similarly in the case of self-interacting systems, where we insist that the two-particle interaction potential vanishes at zero separation.} (hence the sum is starting from $n=1$) and that there are $n$ possibilities how factors of $\partial_J^n$ act on $J$. Due to the derivatives being functional derivatives, below we have an additional Dirac $\delta$-function $\delta_D(z_1''-z_2')$ appearing.

\begingroup
\allowdisplaybreaks
\begin{align}
&    \delta n^{(2)}\left( \vec{x}, z \right) = \frac12 \int \!\frac{\diff^3 k}{(2\pi)^3} \exp\left( i \vec{k} \cdot \vec{x} \right) \!\int \! \diff^3 x^{(i)} \! \int \diff^3 p^{(i)} \, f_0(p^{(i)})  \exp\left( - i \vec{k} \! \cdot \! \frac{\delta}{i \delta \vec{J}_x(z)} \right) \int_{z_i}^{z_f} \! \diff z_1' \! \int_{z_i}^{z_1'} \! \diff z_1'' \nonumber\\*
&  \hspace{2.5cm}    \times \left[ \int_{z_i}^{z_f} \! \diff z_2' \! \int_{z_i}^{z_2'} \! \diff z_2'' \! \frac{m_\nu \, g(z_2',z_2'')}{(1+z_2'') H(z_2'')} \, i\vec{J}_x(z_2') \cdot \frac{\partial \varphi}{\partial \vec{x}} \left(\left( \vec{x}^{(i)} + g(z_2'',z_i) \vec{p}^{(i)} \right),z_2''\right) \right.\nonumber\\*
& \hspace{7cm} \times \left( \frac{m_\nu \, g(z_1',z_1'')}{(1+z_1'') H(z_1'')} \, i\vec{J}_x(z_1') \cdot \frac{\partial \varphi}{\partial \vec{x}} \left( \frac{\delta}{i\,\delta \vec{J}_x(z_1'')} ,z_1''\right) \right) \nonumber\\*
&    \hspace{2.5cm} + 2 \! \int_{z_i}^{z_f} \! \diff z_2' \! \int_{z_i}^{z_2'}  \! \diff z_2'' \! \frac{m_\nu \, g(z_2',z_2'')}{(1+z_2'') H(z_2'')} \, \delta_D(z_1''-z_2') \frac{\partial \varphi}{\partial \vec{x}} \left(\left( \vec{x}^{(i)} + g(z_2'',z_i) \vec{p}^{(i)} \right),z_2''\right) \! \cdot \! \frac{\partial}{\partial \vec{x}} \nonumber\\*
&    \hspace{7cm} \times \left. \left( \frac{m_\nu \, g(z_1',z_1'')}{(1+z_1'') H(z_1'')} \, i\vec{J}_x(z_1') \cdot \frac{\partial \varphi}{\partial \vec{x}} \left( \frac{\delta}{i\,\delta \vec{J}_x(z_1'')} ,z_1''\right) \right)  \right]\nonumber\\*
&    \hspace{5.5cm} \times \restr{\exp \left( i \int_{z_i}^{z_f} \diff z' \, \vec{J}_x(z') \cdot \left( \vec{x}^{(i)} + g(z',z_i) \vec{p}^{(i)} \right) \right)}{\vec{J}_x = 0}
\end{align}
\endgroup
Note the factor of $2$ in the second term. This factor should not be there if the expression above is taken literally. However, the exponential of the interaction operator is actually a time-ordered exponential, as described in a somewhat convoluted way in footnote 3 as well as 4C of \cite{Heisenberg:2022uhb}. Either we use the factor $\frac{1}{2}$ in front, but then allow for both derivatives to act on all factors of $J$ (even to their left) which yields a factor of $2$ in the second term) or we should not have a factor $\frac{1}{2}$ in front at all, but have time-ordering $z_1'<z_2'$ enforced. In both cases this means that ultimately the factor $\frac{1}{2}$ cancels and the final integral boundaries for $z_2'$ are $[z_i,z_1']$.

Now we apply the derivative $\frac{\delta}{i\,\delta \vec{J}_x(z_1'')}$ on the exponential factor according to eq.~\eqref{eq:observable_free_functional}. We also simplify the expression by reordering the bracket,

\begingroup
\allowdisplaybreaks
\begin{align}
     \delta n^{(2)}\left( \vec{x}, z \right) =& \frac12 \int \!\frac{\diff^3 k}{(2\pi)^3} \exp\left( i \vec{k} \cdot \vec{x} \right) \!\int \! \diff^3 x^{(i)} \! \int \diff^3 p^{(i)} \, f_0(p^{(i)})  \exp\left( - i \vec{k} \! \cdot \! \frac{\delta}{i \delta \vec{J}_x(z)} \right) \int_{z_i}^{z_f} \! \diff z_1' \! \int_{z_i}^{z_1'} \! \diff z_1'' \nonumber\\*
    & \times \left[ \int_{z_i}^{z_f} \diff z_2' \int_{z_i}^{z_2'} \diff z_2'' \, \frac{m_\nu \, g(z_2',z_2'')}{(1+z_2'') H(z_2'')} \, i\vec{J}_x(z_2') \cdot \frac{\partial \varphi}{\partial \vec{x}} \left(\left( \vec{x}^{(i)} + g(z_2'',z_i) \vec{p}^{(i)} \right),z_2''\right) \right.\nonumber\\*
    & \left. \quad + 2 \int_{z_i}^{z_f} \diff z_2' \int_{z_i}^{z_2'} \diff z_2'' \, \frac{m_\nu \, g(z_2',z_2'')}{(1+z_2'') H(z_2'')} \, \delta_D(z_1''-z_2') \frac{\partial \varphi}{\partial \vec{x}} \left(\left( \vec{x}^{(i)} + g(z_2'',z_i) \vec{p}^{(i)} \right),z_2''\right) \cdot \frac{\partial}{\partial \vec{x}} \right] \nonumber\\*
    & \hspace{3cm} \times \left( \frac{m_\nu \, g(z_1',z_1'')}{(1+z_1'') H(z_1'')} \, i\vec{J}_x(z_1') \cdot \frac{\partial \varphi}{\partial \vec{x}} \left(\left( \vec{x}^{(i)} + g(z_1'',z_i) \vec{p}^{(i)} \right),z_1''\right) \right)\nonumber\\*
    & \hspace{3cm} \times \restr{\exp \left( i \int_{z_i}^{z_f} \diff z' \, \vec{J}_x(z') \cdot \left( \vec{x}^{(i)} + g(z',z_i) \vec{p}^{(i)} \right) \right)}{\vec{J}_x = 0} 
    \label{eq:number_density_app}
\end{align}   
\endgroup
The next step is to apply the number operator in the first line to the right, i.e. evaluate the functional derivative. This can be done in a straightforward way by applying eq.~\eqref{eq:shift_vector},

\begin{align}
\allowdisplaybreaks
\delta n^{(2)}\left( \vec{x}, z \right) = & \frac12 \int  \frac{\diff^3 k}{(2\pi)^3} \, \exp\left( i \vec{k} \cdot \vec{x} \right) \int \diff^3 x^{(i)} \int \diff^3 p^{(i)} \, f_0(p^{(i)}) \int_{z_i}^{z_f} \diff z_1' \int_{z_i}^{z_1'} \diff z_1''  \nonumber\\*
    & \times \left[ \int_{z_i}^{z_f} \diff z_2' \int_{z_i}^{z_2'} \diff z_2'' \, \frac{m_\nu \, g(z_2',z_2'')}{(1+z_2'') H(z_2'')} \, \delta_D\left( z - z_2' \right) \, (-i\vec{k}) \cdot \frac{\partial \varphi}{\partial \vec{x}} \left(\left( \vec{x}^{(i)} + g(z_2'',z_i) \vec{p}^{(i)} \right),z_2''\right) \right.\nonumber\\*
    &\quad \left. + 2 \int_{z_i}^{z_f} \diff z_2' \int_{z_i}^{z_2'} \diff z_2'' \, \frac{m_\nu \, g(z_2',z_2'')}{(1+z_2'') H(z_2'')} \, \delta_D(z_1''-z_2') \frac{\partial \varphi}{\partial \vec{x}} \left(\left( \vec{x}^{(i)} + g(z_2'',z_i) \vec{p}^{(i)} \right),z_2''\right) \cdot \frac{\partial}{\partial \vec{x}} \right] \nonumber\\*
    & \hspace{1cm} \times  \left( \frac{m_\nu \, g(z_1',z_1'')}{(1+z_1'') H(z_1'')} \, \delta_D\left( z - z_1' \right) \, (-i\vec{k}) \cdot \frac{\partial \varphi}{\partial \vec{x}} \left(\left( \vec{x}^{(i)} + g(z_1'',z_i) \vec{p}^{(i)} \right),z_1''\right) \right)\nonumber\\*
    & \hspace{1.5cm} \times \restr{\exp\left( - i \vec{k} \cdot \frac{\delta}{i \delta \vec{J}_x(z)} \right) \exp \left( i \int_{z_i}^{z_f} \diff z' \, \vec{J}_x(z') \cdot \left( \vec{x}^{(i)} + g(z',z_i) \vec{p}^{(i)} \right) \right)}{\vec{J}_x = 0}  \, .
\end{align}

Note that we already set the terms proportional to the source function $\Vec{J}_x$ from eq.~\eqref{eq:shift_vector} to zero as they will vanish anyhow when $\Vec{J}_x$ is set zero in the end. We can now apply again eq.~\eqref{eq:observable_free_functional} in the last line above and find

\begin{align}
\allowdisplaybreaks
\delta n^{(2)}\left( \vec{x}, z \right) = & \frac12 \int  \frac{\diff^3 k}{(2\pi)^3} \, \exp\left( i \vec{k} \cdot \vec{x} \right) \int \diff^3 x^{(i)} \int \diff^3 p^{(i)} \, f_0(p^{(i)}) \int_{z_i}^{z_f} \diff z_1' \int_{z_i}^{z_1'} \diff z_1''  \nonumber\\*
    & \times \left[ \int_{z_i}^{z_f} \diff z_2' \int_{z_i}^{z_2'} \diff z_2'' \, \frac{m_\nu \, g(z_2',z_2'')}{(1+z_2'') H(z_2'')} \, \delta_D\left( z - z_2' \right) \, (-i\vec{k}) \cdot \frac{\partial \varphi}{\partial \vec{x}} \left(\left( \vec{x}^{(i)} + g(z_2'',z_i) \vec{p}^{(i)} \right),z_2''\right) \right.\nonumber\\*
    &\quad \left. + 2 \int_{z_i}^{z_f} \diff z_2' \int_{z_i}^{z_2'} \diff z_2'' \, \frac{m_\nu \, g(z_2',z_2'')}{(1+z_2'') H(z_2'')} \, \delta_D(z_1''-z_2') \frac{\partial \varphi}{\partial \vec{x}} \left(\left( \vec{x}^{(i)} + g(z_2'',z_i) \vec{p}^{(i)} \right),z_2''\right) \cdot \frac{\partial}{\partial \vec{x}} \right] \nonumber\\*
    & \hspace{1cm} \times  \left( \frac{m_\nu \, g(z_1',z_1'')}{(1+z_1'') H(z_1'')} \, \delta_D\left( z - z_1' \right) \, (-i\vec{k}) \cdot \frac{\partial \varphi}{\partial \vec{x}} \left(\left( \vec{x}^{(i)} + g(z_1'',z_i) \vec{p}^{(i)} \right),z_1''\right) \right)\nonumber\\*
    & \hspace{1.5cm} \qquad \times \exp \left( -i\vec{k} \cdot \left( \vec{x}^{(i)} + g(z',z_i) \vec{p}^{(i)} \right) \right)  \, .
    \label{eq:number_density_app2}
\end{align}
After reordering expression~\eqref{eq:number_density_app2} and performing the redshift integrals $\diff z_1'$ and $\diff z_2'$ we have
\begingroup
\allowdisplaybreaks
\begin{align}
\delta n^{(2)}\left( \vec{x}, z \right) = \frac{1}{2} \int & \diff^3 p^{(i)}  \, f_0( p^{(i)})  \int \diff^3 x^{(i)} \int \frac{\diff^3 k}{(2\pi)^3} \exp\left( - i \vec{k} \cdot \left( \vec{x}^{(i)} + g(z,z_i) \vec{p}^{(i)} - \Vec{x} \right) \right) \nonumber\\*
    & \times \int_{z_i}^{z} \diff z_1'' \, \left[ \int_{z_i}^{z} \diff z_2'' \, \frac{m_\nu \, g(z,z_2'')}{(1+z_2'') H(z_2'')} (-i\vec{k}) \cdot \frac{\partial \varphi}{\partial \vec{x}} \left( \vec{x}^{(i)} + g(z_2'',z_i) \vec{p}^{(i)} ,z_2''\right) \right.\nonumber\\*
    & \hspace{2cm} \left. + 2\int_{z_i}^{z_1''} \diff z_2'' \, \frac{m_\nu \, g(z_1'',z_2'')}{(1+z_2'') H(z_2'')} \, \frac{\partial \varphi}{\partial \vec{x}} \left( \vec{x}^{(i)} + g(z_2'',z_i) \vec{p}^{(i)},z_2''\right) \cdot \frac{\partial}{\partial \vec{x}} \right] \nonumber\\*
    & \times \left( \frac{m_\nu \, g(z,z_1'')}{(1+z_1'') H(z_1'')} (-i\vec{k}) \cdot \frac{\partial \varphi}{\partial \vec{x}} \left( \vec{x}^{(i)} + g(z_1'',z_i) \vec{p}^{(i)} ,z_1''\right) \right) 
\end{align}
\endgroup
We now use the symmetry of the first term in $(z_1'',z_2'')$ to change the integration boundaries of the $z_2''$-integration to $[z_i,z_1'']$ at the cost of an additional factor of $2$. This factor conveniently cancels the prefactor $\frac{1}{2}$ as alluded to earlier,
\begingroup
\allowdisplaybreaks
\begin{align}
\delta n^{(2)}\left( \vec{x}, z \right) = \frac{1}{2} \int & \diff^3 p^{(i)}  \, f_0( p^{(i)})  \int \diff^3 x^{(i)} \int \frac{\diff^3 k}{(2\pi)^3} \exp\left( - i \vec{k} \cdot \left( \vec{x}^{(i)} + g(z,z_i) \vec{p}^{(i)} - \Vec{x} \right) \right) \nonumber\\*
    & \times \int_{z_i}^{z} \diff z_1'' \int_{z_i}^{z_1''} \diff z_2'' \, \left[ \frac{m_\nu \, g(z,z_2'')}{(1+z_2'') H(z_2'')} (-i\vec{k}) \cdot \frac{\partial \varphi}{\partial \vec{x}} \left( \vec{x}^{(i)} + g(z_2'',z_i) \vec{p}^{(i)} ,z_2''\right) \right.\nonumber\\*
    & \hspace{4cm} \left. + \frac{m_\nu \, g(z_1'',z_2'')}{(1+z_2'') H(z_2'')} \, \frac{\partial \varphi}{\partial \vec{x}} \left( \vec{x}^{(i)} + g(z_2'',z_i) \vec{p}^{(i)},z_2''\right) \cdot \frac{\partial}{\partial \vec{x}} \right] \nonumber\\*
    & \times \left( \frac{m_\nu \, g(z,z_1'')}{(1+z_1'') H(z_1'')} (-i\vec{k}) \cdot \frac{\partial \varphi}{\partial \vec{x}} \left( \vec{x}^{(i)} + g(z_1'',z_i) \vec{p}^{(i)} ,z_1''\right) \right) \, .
\end{align}
\endgroup
The next step is the Fourier transform over $k$. Care needs to be taken due to the multiple factors of $\vec{k}$ which do \textit{not} allow for simple replacement by derivatives. In order to proceed with maximal care, we first write the two terms separately and sort them conveniently. Note that in the third line below the scalar products are between the factors of $\vec{k}$ and the factors of $\frac{\partial \varphi}{\partial \vec{x}}$, respectively. For the second term, the derivative $\frac{\partial}{\partial \vec{x}}$ acts on the result of the scalar product in the fourth line, hence we needed to write the scalar product in components (sum over $\mu$ and $\nu$ is implied) in order to keep everything compatible,
\begingroup
\allowdisplaybreaks
\begin{align}
 \delta n^{(2)}\left( \vec{x}, z \right) = \int & \diff^3 p^{(i)}  \, f_0(p^{(i)})  \int \diff^3 x^{(i)} \int_{z_i}^{z} \diff z_1''  \frac{m_\nu \, g(z,z_1'')}{(1+z_1'') H(z_1'')} \int_{z_i}^{z_1''} \diff z_2'' \frac{m_\nu }{(1+z_2'') H(z_2'')} \int \frac{\diff^3 k}{(2\pi)^3} \nonumber\\*
    & \times \Bigg[ g(z,z_2'') \, (-i k_{\mu}) \, (-i k_{\nu}) \exp\left( - i \vec{k}  \left( \vec{x}^{(i)} + g(z,z_i) \vec{p}^{(i)} - \Vec{x} \right) \right)   \nonumber\\*
    & \qquad\quad \times  \frac{\partial \varphi}{\partial x^{\mu}} \left( \vec{x}^{(i)} + g(z_2'',z_i) \vec{p}^{(i)} ,z_2''\right)   \frac{\partial \varphi}{\partial x^{\nu}} \left( \vec{x}^{(i)} + g(z_1'',z_i) \vec{p}^{(i)} ,z_1''\right) \nonumber\\*
    & \quad + g(z_1'',z_2'') \, (-i k_\mu) \exp\left( - i \vec{k} \cdot \left( \vec{x}^{(i)} + g(z,z_i) \vec{p}^{(i)} - \Vec{x} \right) \right) \nonumber\\*
    &\qquad\quad \times  \frac{\partial}{\partial \vec{x}} \frac{\partial \varphi}{\partial x^\mu } \left( \vec{x}^{(i)} + g(z_1'',z_i) \vec{p}^{(i)} ,z_1''\right) \cdot \frac{\partial \varphi}{\partial \vec{x}} \left( \vec{x}^{(i)} + g(z_2'',z_i) \vec{p}^{(i)},z_2''\right) \Bigg] \, . 
\end{align}
\endgroup
We now replace the vectors $\vec{k}$ by derivatives with respect to $\vec{x}^{(i)}$ which exclusively act on the exponential factor following them (as indicated by the brackets). Note that the structure of the scalar products remains unaffected,
\begingroup
\allowdisplaybreaks
\begin{align}
 \delta n^{(2)}\left( \vec{x}, z \right) = \int & \diff^3 p^{(i)}  \, f_0(p^{(i)})  \int \diff^3 x^{(i)} \int_{z_i}^{z} \diff z_1''  \frac{m_\nu \, g(z,z_1'')}{(1+z_1'') H(z_1'')} \int_{z_i}^{z_1''} \diff z_2'' \frac{m_\nu}{(1+z_2'') H(z_2'')} \int \frac{\diff^3 k}{(2\pi)^3}  \nonumber\\*
    &\times  \Bigg( g(z,z_2'') \, \frac{\partial \varphi}{\partial x^{\mu}} \left( \vec{x}^{(i)} + g(z_2'',z_i) \vec{p}^{(i)} ,z_2''\right)   \frac{\partial \varphi}{\partial x^{\nu}} \left( \vec{x}^{(i)} + g(z_1'',z_i) \vec{p}^{(i)} ,z_1''\right)   \nonumber\\*
    & \hspace{3cm} \times \left[ \frac{\partial}{\partial x^{(i)}_{\mu}} \, \frac{\partial}{\partial x^{(i)}_{\nu}} \exp\left( - i \vec{k}  \left( \vec{x}^{(i)} + g(z,z_i) \vec{p}^{(i)} - \Vec{x} \right) \right) \right]  \nonumber\\*
    &\qquad +  g(z_1'',z_2'')  \frac{\partial \varphi}{\partial \vec{x}} \left( \vec{x}^{(i)} + g(z_2'',z_i) \vec{p}^{(i)},z_2''\right) \cdot \frac{\partial}{\partial \vec{x}} \frac{\partial \varphi}{\partial x^\mu} \left( \vec{x}^{(i)} + g(z_1'',z_i) \vec{p}^{(i)} ,z_1''\right) \nonumber \\
    & \hspace{4cm} \times\left[ \frac{\partial}{\partial x^{(i)}_\mu} \exp\left( - i \vec{k} \cdot \left( \vec{x}^{(i)} + g(z,z_i) \vec{p}^{(i)} - \Vec{x} \right) \right) \right] \Bigg) \, .
\end{align}
\endgroup
To shift the derivatives we now perform integration by parts. The surface terms thereby vanish as $\varphi$ and its derivatives are zero at infinity. 
Quite evidently, we need to respect the appearance of product rules, which is the deeper reason for us not having been allowed to simply replace the factors of $\vec{k}$ by derivatives directly. Note that when acting on $\frac{\partial \varphi}{\partial \vec{x}}$, the derivative $\frac{\partial}{\partial \vec{x}^{(i)}}$ can also be written as a derivative with respect to the first argument, denoted by $\frac{\partial}{\partial \vec{x}}$. In the expression below each derivative acts only on the single expression following it (as indicated by the use of brackets). We denote by $\frac{\partial^2 \varphi}{\partial \vec{x}^2}$ the Laplacian of $\varphi$ (with respect to its first argument).

\begingroup
\allowdisplaybreaks
\begin{align}
\delta n^{(2)}\left( \vec{x}, z \right) =  \int & \diff^3 p^{(i)}  \, f_0(p^{(i)})  \int \diff^3 x^{(i)} \int_{z_i}^{z} \diff z_1''  \frac{m_\nu \, g(z,z_1'')}{(1+z_1'') H(z_1'')} \int_{z_i}^{z_1''} \diff z_2'' \frac{m_\nu}{(1+z_2'') H(z_2'')} \, \nonumber\\*
    & \times \Bigg(  g(z,z_2'') \left[ \frac{\partial}{\partial \vec{x}} \frac{\partial^2 \varphi}{\partial \vec{x}^2} \left( \vec{x}^{(i)} + g(z_2'',z_i) \vec{p}^{(i)} ,z_2''\right) \right] \cdot \left[ \frac{\partial \varphi}{\partial \vec{x}} \left( \vec{x}^{(i)} + g(z_1'',z_i) \vec{p}^{(i)} ,z_1''\right) \right] \nonumber\\*
    & + g(z,z_2'') \left[ \frac{\partial \varphi}{\partial \vec{x}} \left( \vec{x}^{(i)} + g(z_2'',z_i) \vec{p}^{(i)} ,z_2''\right) \right] \cdot \left[ \frac{\partial}{\partial \vec{x}} \frac{\partial^2 \varphi}{\partial \vec{x}^2} \left( \vec{x}^{(i)} + g(z_1'',z_i) \vec{p}^{(i)} ,z_1''\right) \right] \nonumber\\*
    & + g(z,z_2'') \left[ \frac{\partial^2 \varphi}{\partial \vec{x}^2} \left( \vec{x}^{(i)} + g(z_2'',z_i) \vec{p}^{(i)} ,z_2''\right) \right] \left[ \frac{\partial^2 \varphi}{\partial \vec{x}^2} \left( \vec{x}^{(i)} + g(z_1'',z_i) \vec{p}^{(i)} ,z_1''\right) \right] \nonumber\\*
    & + g(z,z_2'') \left[ \frac{\partial}{\partial x^\mu} \frac{\partial \varphi}{\partial x^{\nu}} \left( \vec{x}^{(i)} + g(z_2'',z_i) \vec{p}^{(i)} ,z_2''\right) \right] \left[ \frac{\partial}{\partial x_\mu} \frac{\partial \varphi}{\partial x_{\nu}} \left( \vec{x}^{(i)} + g(z_1'',z_i) \vec{p}^{(i)} ,z_1''\right) \right] \nonumber\\*
    & -  g(z_1'',z_2'') \left[ \frac{\partial}{\partial x^\mu} \frac{\partial \varphi}{\partial x^{\nu}} \left( \vec{x}^{(i)} + g(z_2'',z_i) \vec{p}^{(i)},z_2''\right) \right] \left[ \frac{\partial}{\partial x_\mu} \frac{\partial \varphi}{\partial x_\nu} \left( \vec{x}^{(i)} + g(z_1'',z_i) \vec{p}^{(i)} ,z_1''\right) \right] \nonumber\\*
    & - g(z_1'',z_2'') \left[ \frac{\partial \varphi}{\partial \vec{x}} \left( \vec{x}^{(i)} + g(z_2'',z_i) \vec{p}^{(i)},z_2''\right) \right] \cdot \left[ \frac{\partial}{\partial \vec{x}} \frac{\partial^2 \varphi}{\partial \vec{x}^2} \left( \vec{x}^{(i)} + g(z_1'',z_i) \vec{p}^{(i)} ,z_1''\right) \right] \Bigg) \nonumber \\
    & \hspace{2cm} \times \int \frac{\diff^3 k}{(2\pi)^3} \exp\left( - i \vec{k} \cdot \left( \vec{x}^{(i)} + g(z,z_i) \vec{p}^{(i)} - \vec{x} \right) \right) \, .
\end{align}
\endgroup
Finally, the Fourier transform in the last line above simply yields a Dirac $\delta$-function 
of the form $\delta_D\left( \vec{x}^{(i)} + g(z,z_i) \vec{p}^{(i)} - \vec{x} \right)$. The $\mathrm{d}x^{(i)}$ integral can then be performed trivially, where we apply $g(z_1,z_2) = g(z_1,z_i) - g(z_2,z_i)$,

\begingroup
\allowdisplaybreaks
\begin{align}
\delta n^{(2)}\left( \vec{x}, z \right) =  \int & \diff^3 p^{(i)}  \, f_0(p^{(i)}) \int_{z_i}^{z} \diff z_1''  \frac{m_\nu \, g(z,z_1'')}{(1+z_1'') H(z_1'')} \int_{z_i}^{z_1''} \diff z_2'' \frac{m_\nu}{(1+z_2'') H(z_2'')} \, \nonumber\\*
    & \times \Bigg(  g(z,z_2'') \left[ \frac{\partial}{\partial \vec{x}} \frac{\partial^2 \varphi}{\partial \vec{x}^2} \left( \vec{x}^{(i)} - g(z, z_2'') \vec{p}^{(i)} ,z_2''\right) \right] \cdot \left[ \frac{\partial \varphi}{\partial \vec{x}} \left( \vec{x}^{(i)} - g(z,z_1'') \vec{p}^{(i)} ,z_1''\right) \right] \nonumber\\*
    & + g(z,z_2'') \left[ \frac{\partial \varphi}{\partial \vec{x}} \left( \vec{x} - g(z,z_2'') \vec{p}^{(i)} ,z_2''\right) \right] \cdot \left[ \frac{\partial}{\partial \vec{x}} \frac{\partial^2 \varphi}{\partial \vec{x}^2} \left( \vec{x} - g(z,z_1'') \vec{p}^{(i)} ,z_1''\right) \right] \nonumber\\*
    & + g(z,z_2'') \left[ \frac{\partial^2 \varphi}{\partial \vec{x}^2} \left( \vec{x} - g(z,z_2'') \vec{p}^{(i)} ,z_2''\right) \right] \left[ \frac{\partial^2 \varphi}{\partial \vec{x}^2} \left( \vec{x} - g(z,z_1'') \vec{p}^{(i)} ,z_1''\right) \right] \nonumber\\*
    & + g(z,z_2'') \left[ \frac{\partial}{\partial x^\mu} \frac{\partial \varphi}{\partial x^{\nu}} \left( \vec{x} - g(z,z_2'') \vec{p}^{(i)} ,z_2''\right) \right] \left[ \frac{\partial}{\partial x_\nu} \frac{\partial \varphi}{\partial x_{\mu}} \left( \vec{x} - g(z,z_1'') \vec{p}^{(i)} ,z_1''\right) \right] \nonumber\\*
    & -  g(z_1'',z_2'') \left[ \frac{\partial}{\partial x^\mu} \frac{\partial \varphi}{\partial x^{\nu}} \left( \vec{x} - g(z,z_2'') \vec{p}^{(i)},z_2''\right) \right] \left[ \frac{\partial}{\partial x_\mu} \frac{\partial \varphi}{\partial x_\nu} \left( \vec{x} - g(z,z_1'') \vec{p}^{(i)} ,z_1''\right) \right] \nonumber\\*
    & - g(z_1'',z_2'') \left[ \frac{\partial \varphi}{\partial \vec{x}} \left( \vec{x} - g(z,z_2'') \vec{p}^{(i)},z_2''\right) \right] \cdot \left[ \frac{\partial}{\partial \vec{x}} \frac{\partial^2 \varphi}{\partial \vec{x}^2} \left( \vec{x} - g(z,z_1'') \vec{p}^{(i)} ,z_1''\right) \right]  \Bigg) \, .
\end{align}
\endgroup

We can now combine the second with the sixth term as well as the fourth with the fifth term by using $g(z_1,z_2) = g(z,z_2) - g(z,z_1)$. The final expression is given in eq.~\eqref{eq:KFT_2nd_order} in the main text.

\section{Second-order Vlasov derivation}
\label{app:Second-order Vlasov derivation}

Our starting point is again the Vlasov equation in Fourier-space in eq.~\eqref{eq:Vlasov_Fourier}. As we already derived its solution at linear order in the main text in sec.~\ref{sec:Perturbative solutions to the Vlasov equation}, we here focus on the derivation of the second-order solution. At second order, the Vlasov equation in Fourier-space is

\begin{align}
\frac{\partial}{\partial s} \delta \tilde{f}^{(2)} (\Vec{k}) + i \frac{\Vec{k}\cdot \Vec{p}}{m_{\nu}} \delta \Tilde{f}^{(2)}(\Vec{k}) - i a^2 m_{\nu} \int \frac{\mathrm{d}^3k_1}{(2\pi)^3} \frac{\mathrm{d}^3 k_2}{(2\pi)^3} \, (2 \pi)^3 \delta(\Vec{k}-\Vec{k}_1-\Vec{k}_2) \, \Vec{k}_1 \cdot \frac{\partial \delta \Tilde{f}^{(1)} (\Vec{k}_2)}{\partial \Vec{p}} \, \Tilde{\varphi}(\Vec{k}_1) = 0
\end{align}
The integral solution to this can be obtained analogously to eq.~\eqref{eq:trick1} and is given by
\begin{align}
  \delta \Tilde{f}^{(2)}(\Vec{k})  =  i m_{\nu} \! \int_{s_i}^s \!\mathrm{d}s' \, a^2(s') \int \! \frac{\mathrm{d}^3 k_1}{(2\pi)^3} \int \! \frac{\mathrm{d}^3 k_2}{(2\pi)^3} \, (2\pi)^3 \delta(\Vec{k}-\Vec{k}_1-\Vec{k}_2) \, \Vec{k}_1 \cdot \frac{\partial \delta \Tilde{f}^{(1)}(\Vec{k}_2)}{\partial \Vec{p}} \, \Tilde{\varphi}(\Vec{k_1}) \, e^{i \frac{\Vec{k}\cdot\Vec{p}}{m_{\nu}}(s'-s)} \, .
    \label{eq:deltaf_Vlasov}
\end{align}

To obtain the number density we now integrate the above expression over momentum. Following the derivation at linear order, we thereby integrate by parts and apply the divergence theorem (see eq.~\eqref{eq:number_Vlasov_1}) and find
\begin{align}
   \delta \Tilde{n}^{(2)}(\Vec{k}) \! = \! \! \int_{s_i}^s \! \! \mathrm{d}s' a^2(s') \, (s' \! - \!s) \! \! \int \! \!\frac{\mathrm{d}^3k_1}{(2\pi)^3} \! \! \int \! \! \frac{\mathrm{d}^3 k_2}{(2\pi)^3}  (2\pi)^3 \delta(\Vec{k} \! - \! \Vec{k}_1 \! - \! \Vec{k}_2) \left( \Vec{k}_1 \cdot \Vec{k} \right)  \Tilde{\varphi}(\Vec{k}_1) \! \int \! \mathrm{d}^3 p \, \delta \Tilde{f}^{(1)}(\Vec{k}_2,s') e^{i \frac{\Vec{k} \! \cdot \! \Vec{p}}{m_{\nu}}(s' \! - \! s)} \, .
    \label{eq:n_Vlasov}
\end{align}
We now proceed by inserting the first-order solution for $\delta \Tilde{f}^{(1)}$ from eq.~\eqref{eq:delta_f1_sol} into the above expression,

\begin{align}
  \delta \Tilde{n}^{(2)} (\Vec{k}) &= i m_{\nu} \int_{s_i}^s \mathrm{d}s' \, a^2(s') (s'-s) \! \int \frac{\mathrm{d}^3k_1}{(2\pi)^3} \! \int \frac{\mathrm{d}^3k_2}{(2\pi)^3} \, (2\pi)^3 \delta(\Vec{k}-\Vec{k}_1-\Vec{k}_2) \, \left( \Vec{k}_1 \cdot \Vec{k} \right) \Tilde{\varphi}(\Vec{k}_1,s') \\
    & \hspace{3cm} \times \int_{s_i}^{s'} \mathrm{d}s'' \, a^2(s'') \, \Tilde{\varphi}(\Vec{k}_2,s'') \, \Vec{k}_2 \cdot \int \mathrm{d}^3 p \, \frac{\partial f_0}{\partial \Vec{p}} e^{i \frac{\Vec{p}}{m_{\nu}}\left( \Vec{k}(s'-s)+\Vec{k}_2(s''-s) \right)}
\end{align}

We again apply partial integration and the divergence theorem for the momentum integration (analogously to eq.~\eqref{eq:number_Vlasov_1} at linear order),

\begin{align}
   \delta n^{(2)} (\Vec{k}) =& \int_{s_i}^s \mathrm{d}s' \, a^2(s') (s'-s)^2 \int_{s_i}^{s'} \mathrm{d}s'' \, a^2(s'') \! \int \frac{\mathrm{d}^3k_1}{(2\pi)^3} \! \int \frac{\mathrm{d}^3k_2}{(2\pi)^3} \, \delta(\Vec{k}-\Vec{k}_1-\Vec{k}_2) \, \left( \Vec{k}_1 \! \cdot \! \Vec{k} \right) \left( \Vec{k}_2 \! \cdot \! \Vec{k} \right) \, \nonumber \\
   & \hspace{4cm} \times (2\pi)^3 \Tilde{\varphi}(\Vec{k}_1,s') \Tilde{\varphi}(\Vec{k}_2,s'') \int \mathrm{d}^3p \, e^{i \frac{\Vec{p}\cdot \Vec{k}}{m_{\nu}}(s'-s)} e^{i \frac{\Vec{p}\cdot\Vec{k}_2}{m_{\nu}}(s''-s')} \nonumber \\
   & + \int_{s_i}^s \! \mathrm{d}s' \, a^2(s') (s'-s) \! \int_{s_i}^{s'} \! \mathrm{d}s'' \, a^2(s'') (s''-s') \! \int \frac{\mathrm{d}^3 k_1}{(2\pi)^3} \int \frac{\mathrm{d}^3k_2}{(2\pi)^3} \, \delta(\Vec{k}-\Vec{k}_1-\Vec{k}_2) \left( \Vec{k}_1 \! \cdot \! \Vec{k} \right) k_2^2 \nonumber \\
   & \hspace{4cm} \times (2\pi)^3 \Tilde{\varphi}(\Vec{k}_1,s') \Tilde{\varphi}(\Vec{k}_2,s'') \int \mathrm{d}^3p \, e^{i \frac{\Vec{p}\cdot \Vec{k}}{m_{\nu}}(s'-s)} e^{i \frac{\Vec{p}\cdot\Vec{k}_2}{m_{\nu}}(s''-s')} \, .
\end{align}

After Fourier transformation and re-ordering of the above expression we find
\begin{align}
    \delta n^{(2)} (\Vec{x}) &=  \int_{s_i}^s \mathrm{d}s' \, a^2(s') (s' \! - \! s)^2 \! \int_{s_i}^{s'} \mathrm{d}s'' a^2(s'') \! \int \mathrm{d}^3p \, f_0(p) \! \int \frac{\mathrm{d}^3k_1}{(2\pi)^3} \! \int \frac{\mathrm{d}^3k_2}{(2\pi)^3} \, \Tilde{\varphi}(\Vec{k_1},s') \Tilde{\varphi}(\Vec{k}_2,s'') \nonumber \\
    & \quad \times e^{i \frac{\Vec{p}\cdot \Vec{k}_2}{m_{\nu}}(s''-s')} \int \mathrm{d}^3k \, \delta(\Vec{k}-\Vec{k_1}-\Vec{k_2}) \left( \Vec{k}_1 \cdot \Vec{k} \right) \left( \Vec{k}_2 \cdot \Vec{k} \right) e^{i \Vec{k}\cdot\Vec{x} + i \frac{\Vec{p}\cdot \Vec{k}}{m_{\nu}}(s'-s)} \nonumber \\
    & + \int_{s_i}^s \! \mathrm{d}s' \, a^2(s') (s' \! - \! s) \! \int_{s_i}^{s'}\! \mathrm{d}s'' a^2(s'') (s'' \! - \! s') \! \int \mathrm{d}^3p \, f_0(p) \! \int \frac{\mathrm{d}^3k_1}{(2\pi)^3} \! \int \frac{\mathrm{d}^3k_2}{(2\pi)^3} \, \Tilde{\varphi}(\Vec{k_1},s') \Tilde{\varphi}(\Vec{k}_2,s'') \nonumber \\
    & \quad \times e^{i \frac{\Vec{p}\cdot \Vec{k}_2}{m_{\nu}}(s''-s')} \int \mathrm{d}^3k \, \delta(\Vec{k}-\Vec{k_1}-\Vec{k_2}) \left( \Vec{k}_1 \cdot \Vec{k} \right) k_2^2 \, e^{i \Vec{k}\cdot\Vec{x} + i \frac{\Vec{p}\cdot \Vec{k}}{m_{\nu}}(s'-s)} \\
    =& \int_{s_i}^s \mathrm{d}s' \, a^2(s') (s' \! - \! s)^2 \! \int_{s_i}^{s'} \mathrm{d}s'' a^2(s'') \! \int \mathrm{d}^3p \, f_0(p) \! \int \frac{\mathrm{d}^3k_1}{(2\pi)^3} \! \int \frac{\mathrm{d}^3k_2}{(2\pi)^3} \nonumber \\
    & \quad \times \Tilde{\varphi}(\Vec{k}_1,s') \Tilde{\varphi}(\Vec{k}_2,s'') \left( \Vec{k}_1 \cdot \left( \Vec{k}_1 + \Vec{k}_2 \right)\right) \left( \Vec{k}_2 \cdot \left( \Vec{k}_1 + \Vec{k}_2 \right) \right) e^{i \Vec{k}_1\cdot\Vec{x} + i \frac{\Vec{p}\cdot\Vec{k}_1}{m_{\nu}}(s'-s)} e^{i \Vec{k}_2\cdot\Vec{x} + i \frac{\Vec{p}\cdot\Vec{k}_2}{m_{\nu}}(s''-s)} \nonumber \\
    & + \int_{s_i}^s \! \mathrm{d}s' \, a^2(s') (s' \! - \! s) \! \int_{s_i}^{s'} \mathrm{d}s'' a^2(s'') (s'' \! - \! s') \! \int \mathrm{d}^3p \, f_0(p) \! \int \frac{\mathrm{d}^3k_1}{(2\pi)^3} \! \int \frac{\mathrm{d}^3k_2}{(2\pi)^3} \nonumber \\
    & \quad \times \Tilde{\varphi}(\Vec{k}_1,s') \Tilde{\varphi}(\Vec{k}_2,s'') \left( \Vec{k}_1 \cdot \left( \Vec{k}_1 + \Vec{k}_2 \right)\right) k_2^2 \,  e^{i \Vec{k}_1\cdot\Vec{x} + i \frac{\Vec{p}\cdot\Vec{k}_1}{m_{\nu}}(s'-s)} e^{i \Vec{k}_2\cdot\Vec{x} + i \frac{\Vec{p}\cdot\Vec{k}_2}{m_{\nu}}(s''-s)} \, .
\end{align} 

Writing out the scalar products as
\begin{align}
    \left( \Vec{k}_1 \cdot \left( \Vec{k}_1+\Vec{k}_2 \right) \right) \left( \Vec{k}_2 \cdot \left( \Vec{k}_1 + \Vec{k}_2 \right)\right) &= k_1^2\left( \Vec{k}_1 \cdot \Vec{k}_2 \right) + \left( \Vec{k}_1 \cdot \Vec{k}_2 \right) \left( \Vec{k}_1 \cdot \Vec{k}_2 \right) + k_2^2 \left( \Vec{k}_1 \cdot \Vec{k}_2 \right) + k_1^2 k_2^2 \, , \\
    \left( \Vec{k}_1 \cdot \left( \Vec{k}_1+ \Vec{k}_2 \right) \right) k_2^2 &= k_1^2 k_2^2 + \left( \Vec{k}_1 \cdot \Vec{k}_2 \right) k_2^2  
\end{align}
shows that we end up with six terms,
\begin{align}
    \delta n^{(2)}(\Vec{x}) =& \int \mathrm{d}^3p \, f_0(p) \int_{s_i}^s \mathrm{d}s' \, a^2(s') (s'-s) \int_{s_i}^{s'} \mathrm{d}s'' \, a^2(s'') \nonumber \\
     \times & \left( (s'-s) \left[ \! \int \! \frac{\mathrm{d}^3k_1}{(2\pi)^3} \, k_1^2 \Vec{k}_1 \, \Tilde{\varphi}(\Vec{k}_1,s') e^{i \Vec{k}_1 \cdot \Vec{x} +i \frac{\Vec{p}\cdot \Vec{k}_1}{m_{\nu}}(s'-s)} \right] \cdot \left[ \! \int \! \frac{\mathrm{d}^3k_2}{(2\pi)^3} \, \Vec{k}_2 \, \Tilde{\varphi}(\Vec{k}_2,s'') e^{i \Vec{k}_2\cdot\Vec{x} + i \frac{\Vec{p} \cdot \Vec{k}_2}{m_{\nu}} (s''-s)} \right] \right. \nonumber \\ 
     & \left. + (s'-s) \left[ \! \int \! \frac{\mathrm{d}^3k_1}{(2\pi)^3} \, k_1^{\mu} k_2^{\nu} \, \Tilde{\varphi}(\Vec{k}_1,s') e^{i \Vec{k}_1 \cdot \Vec{x} +i \frac{\Vec{p}\cdot \Vec{k}_1}{m_{\nu}}(s'-s)} \right] \cdot \left[ \! \int \! \frac{\mathrm{d}^3k_2}{(2\pi)^3} \, k_{1 \mu} k_{2 \nu} \, \Tilde{\varphi}(\Vec{k}_2,s'') e^{i \Vec{k}_2\cdot\Vec{x} + i \frac{\Vec{p} \cdot \Vec{k}_2}{m_{\nu}} (s''-s)} \right] \right. \nonumber \\
     & \left. + (s'-s) \left[ \! \int \! \frac{\mathrm{d}^3k_1}{(2\pi)^3} \, \Vec{k}_1 \, \Tilde{\varphi}(\Vec{k}_1,s') e^{i \Vec{k}_1 \cdot \Vec{x} +i \frac{\Vec{p}\cdot \Vec{k}_1}{m_{\nu}}(s'-s)} \right] \cdot \left[ \! \int \! \frac{\mathrm{d}^3k_2}{(2\pi)^3} \, \Vec{k}_2 k_2^2 \, \Tilde{\varphi}(\Vec{k}_2,s'') e^{i \Vec{k}_2\cdot\Vec{x} + i \frac{\Vec{p} \cdot \Vec{k}_2}{m_{\nu}} (s''-s)} \right] \right. \nonumber \\
     & \left. + (s'-s) \left[ \! \int \! \mathrm{d}^3k_1 \, k_1^2 \, \Tilde{\varphi}(\Vec{k}_1,s') e^{i \Vec{k}_1 \cdot \Vec{x} +i \frac{\Vec{p}\cdot \Vec{k}_1}{m_{\nu}}(s'-s)} \right] \cdot \left[ \! \int \!  \frac{\mathrm{d}^3k_2}{(2\pi)^3} \, k_2^2 \, \Tilde{\varphi}(\Vec{k}_2,s'') e^{i \Vec{k}_2\cdot\Vec{x} + i \frac{\Vec{p} \cdot \Vec{k}_2}{m_{\nu}} (s''-s)} \right] \right. \nonumber \\
     & \left. + (s''-s') \left[ \! \int \! \frac{\mathrm{d}^3k_1}{(2\pi)^3} \, k_1^2 \, \Tilde{\varphi}(\Vec{k}_1,s') e^{i \Vec{k}_1 \cdot \Vec{x} +i \frac{\Vec{p}\cdot \Vec{k}_1}{m_{\nu}}(s'-s)} \right] \cdot \left[ \! \int \! \frac{\mathrm{d}^3k_2}{(2\pi)^3} \, k_2^2 \, \Tilde{\varphi}(\Vec{k}_2,s'') e^{i \Vec{k}_2\cdot\Vec{x} + i \frac{\Vec{p} \cdot \Vec{k}_2}{m_{\nu}} (s''-s)} \right] \right. \nonumber \\
     & \left. + (s''-s') \left[ \! \int \! \frac{\mathrm{d}^3k_1}{(2\pi)^3} \, \Vec{k}_1 \, \Tilde{\varphi}(\Vec{k}_1,s') e^{i \Vec{k}_1 \cdot \Vec{x} +i \frac{\Vec{p}\cdot \Vec{k}_1}{m_{\nu}}(s'-s)} \right] \cdot \left[ \! \int \! \frac{\mathrm{d}^3k_2}{(2\pi)^3} \, \Vec{k}_2 k_2^2 \, \Tilde{\varphi}(\Vec{k}_2,s'') e^{i \Vec{k}_2\cdot\Vec{x} + i \frac{\Vec{p} \cdot \Vec{k}_2}{m_{\nu}} (s''-s)} \right] \right) \, .
\end{align}

From the six terms above we can combine the third with the sixth term and the fourth with the fifth term. Replacing the wave vectors with derivatives and performing the Fourier transformations we end up with eq.~\eqref{eq:KFT_2nd_order_s}. 

\section{Model of the gravitational environment}
\label{app:Gravitational environment}

In this section, we describe the gravitational environment of the clustering analysis conducted in this paper. As discussed in sec. \ref{sec:calculation}, in this work we focus on a spherically symmetric NFW potential (i.e. we neglect all contributions to the gravitational potential other than CDM). In order to facilitate the comparison with state-of-the-art relic neutrino clustering calculations, we model gravitational potential of CDM as in~\cite{Mertsch:2019qjv}. The NFW mass density profile at distances $r<R_\mathrm{vir}$ is \cite{Navarro:1995iw}
\begin{align}
    \rho_\mathrm{NFW} (r,z) = \frac{\rho_0(z)}{\frac{r}{R_s(z)} \left( 1 + \frac{r}{R_s(z)}\right)^2} \,, \hspace{1cm} (r<R_\mathrm{vir} (z))
\end{align}
where $\rho_0(z)$ is a normalization parameter and the virial radius $R_\mathrm{vir}(z)$ is related to the characteristic inner radius $R_s(z)$ through the concentration parameter $R_\mathrm{vir}(z) = c(z) R_s(z)$. We follow \cite{Mertsch:2019qjv,deSalas:2017wtt} and assume for the time dependence of the concentration parameter
\begin{align}
    \log_{10} \left( \beta c(z) \right)= a(z) + b(z) \log_{10} \left( \frac{M_\mathrm{vir}}{10^{12} h^{-1} M_\odot} \right)
\end{align}
with
\begin{align}
    \beta = 0.613 \,, \quad a(z) = 0.537 + (1.025 - 0.537) \mathrm{e}^{-0.718z^{1.08}} \,, \quad b(z) = -0.097 + 0.024z \,.
\end{align}
According to \cite{Mertsch:2019qjv,deSalas:2017wtt} the virial mass $M_\mathrm{vir}$ is assumed to be constant in time,
\begin{equation}
    M_\mathrm{vir} = 2.03 \times 10^{12} M_\odot \,.
\end{equation}
In practice, this amounts to replacing the static values for the scale radius $R_s$ and the virial radius $R_\mathrm{vir}$ by their time-dependent analogues given by

\begin{align}
    R_\mathrm{vir} (z)= c(z) R_s (z) = \left(\frac{3 M_\mathrm{vir}}{4\pi \Delta_\mathrm{vir} (z) \rho_\mathrm{crit} (z) } \right)^{1/3} \,, 
    \label{eq:scale_radius}
\end{align}
with
\begin{equation}
     \Delta_\mathrm{vir} (z) = 18\pi^2 + 82 (\Omega_m (z) - 1) - 39 (\Omega_m (z) - 1)^2 \,, \qquad \rho_\mathrm{crit}(z) = 3 H(z)^2 /(8\pi G)
\end{equation}
and the fractional matter density is given by
\begin{equation}
    \Omega_m(z)=\frac{\Omega_{0,m} (1 + z)^3}{\Omega_{0,m} (1 + z)^3 + \Omega_{0,\Lambda}}.
\end{equation}
The resulting gravitational potential can be derived from solving the Poisson equation. Following \cite{deSalas:2017wtt,Mertsch:2019qjv}, outside its virial radius the gravitational potential is simply assumed to be the one of a point source with mass $M_{\mathrm{vir}}$, such that we have
\begin{align}
\varphi_\mathrm{NFW} (r,z) = \begin{cases}
- \frac{4 \pi G \rho_0(z) R_s(z)^2}{(1+z)^2} \left[ \frac{\ln(1 + r(z)/R_s(z))}{r/R_s(z)} - \frac{1}{1 + R_\mathrm{vir}(z)/R_s(z)} \right] & (r<R_\mathrm{vir} (z))\\
- G  (1+z) \frac{M_{\mathrm{vir}}}{r} & (r>R_\mathrm{vir} (z)) .
    \end{cases}
\end{align}

The time-dependence of the central density $\rho_0(z)$ can be derived from demanding the time-independence of the virial mass,
\begin{equation}
    M_{\mathrm{vir}}= \frac{4 \pi}{(1+z)^3} \int_{0}^{R_{\mathrm{vir}(z)}} \mathrm{d}x \, x^2 \rho_{\mathrm{NFW}}(x,z)
\end{equation}
which finally gives 
\begin{equation}
\rho_0(z) = M_{\mathrm{vir}}\left[\frac{4 \pi}{(1 + z)^3}
   R_s(z)^3 \left(\ln \left( 1+c(z) \right)- \frac{c(z)}{1 + c(z)}  \right) 
      \right]^{-1} \,.
\end{equation}
We emphasize the spherical symmetry of the NFW profile and we make use of this property in the following. The first and second order kinetic field theoretic perturbative corrections to the local relic neutrino number density require the first, second and third derivatives of the gravitational potential, which are contracted and/or multiplied in different ways. 

Explicitly, the four derivative terms of the potential in~\eqref{eq:KFT_2nd_order}
\begin{align}
    \partial_i \partial^i \varphi_\mathrm{NFW}(r_2,z_2) &\partial_j \partial^j \varphi_\mathrm{NFW}(r_1,z_1) = \frac{4\pi G \rho_0(z_2) R_s(z_2)^3}{r_2(r_2 + R_s(z_2))^2 {(1+z_1)^2(1+z_2)^2} } \frac{4\pi G \rho_0(z_1) R_s(z_1)^3}{r_1(r_1 + R_s(z_1))^2} \,,\\
    \partial^i \partial_j \partial^j \varphi_\mathrm{NFW}(r_2,z_2) &\partial_i \varphi_\mathrm{NFW}(r_1,z_1) = (x_2 \cdot x_1) \frac{4\pi G \rho_0(z_2) R_s(z_2)^3 (3 r_2 +R_s(z_2))}{r_2^3(r_2 + R_s(z_2))^3 {(1+z_1)^2(1+z_2)^2}} \label{cross_diff}\, \\
    &\qquad\frac{4\pi G \rho_0(z_1) R_s(z_1)^3}{r_1^3} \left[ \frac{r_1}{r_1 + R_s(z_1)} - \ln \left( 1 + \frac{r_1}{R_s(z_1)} \right) \right]\,,\nonumber \\
    \partial_j \partial^i \varphi_\mathrm{NFW}(r_2,z_2) &\partial_i \partial^j \varphi_\mathrm{NFW}(r_1,z_1) = r_1 R_s(z_1) (2\log(1 + r_1/R_s(z_1)) - 1)\nonumber \\
& \quad \cdot \bigg( r_1^2 r_2^4 (3\log(1 + r_2/R_s(z_2)) - 2)\nonumber \\
& \quad \quad + 3r_2 r_2 \vec{r_1}\cdot \vec{r_2} (3\log(1 + r_2/R_s(z_2)) - 4) \nonumber\\
& \quad \quad + 3r_2 R_s(z_2) (2\log(1 + r_2/R_s(z_2)) - 1) \big(r_1 r_2^2 + 3\vec{r_1}\cdot \vec{r_2}\big) \nonumber\\
& \quad \quad + 3\log(1 + r_2/R_s(z_2)) R_s(z_2)^2 \big(r_1 r_2^2 + 3\vec{r_1}\cdot \vec{r_2}\big) \bigg)\nonumber \\
& + R_s(z_1)^2 \log(1 + r_1/R_s(z_1))\nonumber \\
& \quad \cdot \bigg( r_1^2 r_2^4 (3\log(1 + r_2/R_s(z_2)) - 2) \nonumber\\
& \quad \quad + 3r_2 r_2 \vec{r_1}\cdot \vec{r_2} (3\log(1 + r_2/R_s(z_2)) - 4) \nonumber\\
& \quad \quad + 3r_2 R_s(z_2) (2\log(1 + r_2/R_s(z_2)) - 1) \big(r_1 r_2^2 + 3\vec{r_1}\cdot \vec{r_2}\big) \nonumber\\
& \quad \quad + 3\log(1 + r_2/R_s(z_2)) R_s(z_2)^2 \big(r_1 r_2^2 + 3\vec{r_1}\cdot \vec{r_2}\big) \bigg) \nonumber\\
 + r_1^2 &\bigg( r_1^2 r_2^4 \big((1 - 2\log(1 + r_2/R_s(z_2))) \nonumber\\
& \quad + \log(1 + r_1/R_s(z_1))(3\log(1 + r_2/R_s(z_2)) - 2)\big) \nonumber\\
& \quad + r_2^2 \vec{r_1}\cdot \vec{r_2} (3\log(1 + r_1/R_s(z_1)) - 4) (4\log(1 + r_2/R_s(z_2)) - 4) \nonumber\\
& \quad + r_2 R_s(z_2) (2\log(1 + r_2/R_s(z_2)) - 1) \big(r_1 r_2^2 (3\log(1 + r_1/R_s(z_1)) - 2)\nonumber \\
& \quad + 3\vec{r_1}\cdot \vec{r_2} (3\log(1 + r_1/R_s(z_1)) - 4)\big) \nonumber\\
& \quad + R_s(z_2)^2 \log(1 + r_2/R_s(z_2)) \nonumber\\
& \quad \cdot \big(r_1 r_2^2 (3\log(1 + r_1/R_s(z_1)) - 2) + 3\vec{r_1}\cdot \vec{r_2} (3\log(1 + r_1/R_s(z_1)) - 4)\big) \bigg) .\nonumber
\end{align}

\begin{figure}[tb]
    \centering
    \includegraphics[width=\columnwidth]{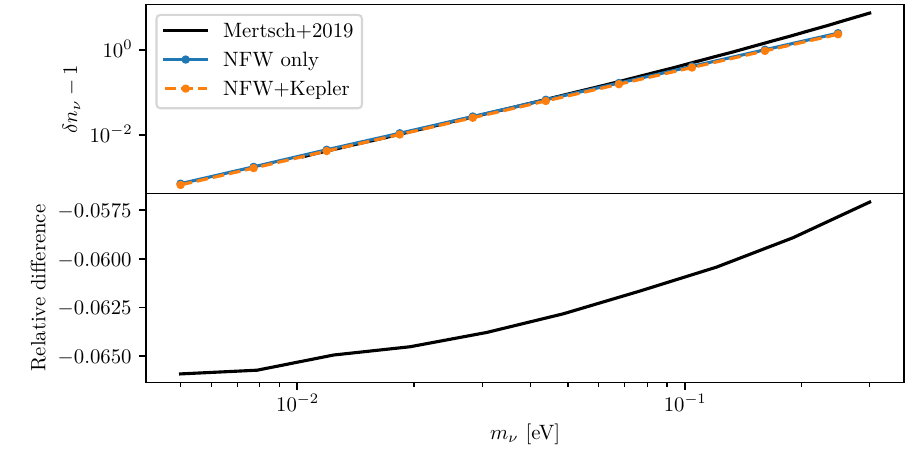}
    \caption{\label{fig:kepler_comparison} \textit{Top panel}: Relic neutrino density as estimated by the backtracking method of~\cite{Mertsch:2019qjv} (black) and in the first-order perturbation theory described in this paper, assuming either a full NFW halo or a transition to a Kepler potential at the virial radius of the NFW halo, as done in~\cite{Mertsch:2019qjv}. \textit{Bottom panel}: Relative difference between the first order prediction with and without the Kepler potential transition. Omitting the Kepler potential results in around $5$--$6 \%$ overestimation of the density.}
\end{figure}
The only difference between our potential and the one used in~\cite{Mertsch:2019qjv} is that the latter assume a Kepler potential at radii $r> R_\mathrm{vir} (z)$. This would heavily complicate the derivatives above. To test its impact, we have implemented the transition to the Kepler potential at the first order. Figure~\ref{fig:kepler_comparison} shows the overdensity at first order, predicted by our code, when assuming only the NFW potential as well as when including the transition to the Kepler potential. Evidently, the transition has only a minor impact, so we do not expect it to affect the conclusions made in the paper.

\section{Numerical details}
\label{app:numerical}

In this appendix, we briefly describe our numerical approach to computing the integral~\eqref{eq:numerical_1}. Firstly, the first- and second-order contributions $\varepsilon^{(1)} (\vec{x}, \vec{p}^{(i)})$ and $\varepsilon^{(2)} (\vec{x}, \vec{p}^{(i)})$ are evaluated with an argument of the form $|\vec{x} - g(z, z') \vec{p}^{(i)}|$, with $\vec{p}^{(i)}$ the integration variable, which we write using spherical coordinates as $\vec{p}^{(i)}=(p^{(i)} \sin \theta \cos \phi, p^{(i)} \sin \theta \sin \phi, p^{(i)} \cos \theta)$. Since we compute the integral in spherical coordinates and $\vec{x}$ is the only quantity defining a fixed direction in space, we will choose a coordinate system in which $\vec{x} = (0, 0, x_2)$, leading to the simplification
\begin{align}
    \left| \vec{x} - g(z, z') \vec{p}^{(i)} \right|^2 &= \left(g(z,z') p^{(i)} \sin \theta \cos \phi\right)^2 + \left(g(z,z') p^{(i)} \sin \theta \sin \phi\right)^2 + \left(x_2 - g(z,z') p^{(i)} \cos \theta\right)^2 \nonumber \\
    &= x_2^2 + g(z,z')^2 {p^{(i)}}^2 - 2 g(z,z') p^{(i)} x_2 \cos \theta. \label{cross_radius}
\end{align}
Crucially, this is independent of the $\phi$ component of the integration variable $p^{(i)}$, thus reducing the dimensionality of the perturbation theory integrals by one. Therefore, we can compute the integral as
\begin{align}
    n (x_2) =  2\pi \int  \mathrm{d} \theta \sin(\theta)  \int  \mathrm{d} y \ T_0^3 f_0(y)  \left[ 1 + \varepsilon^{(1)} (x_2, \theta, y) +  \varepsilon^{(2)} (x_2, \theta, y) \right] ,
\end{align}
where we have substituted $y\equiv p^{(i)}/T_0$. All integrals, including the redshift integrals implicit in the $\varepsilon^{(i)}$ quantities, are carried out using adaptive Gauss-Kronrod quadrature, noting that we cannot employ the stronger Gauss-Laguerre quadrature for the momentum integral due to the cut-off at $y_\mathrm{cut}$~\cite{Press2007}. The propagator~\eqref{eq:defn_propagator} involves evaluating the Gaussian hypergeometric function, whose numerical evaluation is usually slow. To accommodate this, we approximate it with a $(12, 12)$ minimax rational approximation~\cite{Press2007}, as in~\cite{Holm:2023rml}. Ultimately, the first order computation takes on the order of milliseconds, and the second order computation takes on the order of $10^0$--$10^1$ seconds, strongly depending, of course, on the exact parameters such as the neutrino mass and the momentum cut-off. The calculation is coded in \textsc{C++}, borrowing some functions from the \textsc{CLASS} code~\cite{Blas:2011rf, Lesgourgues:2011rh}, and is publicly available at \url{https://github.com/EBHolm/KFT-Neutrinos} on the branch with the arXiv ID of this paper.

\bibliographystyle{utcaps}
\bibliography{Literature}



\end{document}